\documentclass[%
 aip,
 amsmath,amssymb,
 reprint,%
]{revtex4-1}

\usepackage{graphicx}
\usepackage{dcolumn}
\usepackage{bm}
\usepackage{cleveref}

\usepackage{todonotes}

\crefname{section}{Sec.}{Secs.}
\Crefname{section}{Section}{Sections}
\crefname{subsection}{Sec.}{Secs.}
\Crefname{subsection}{Section}{Sections}
\crefname{figure}{Fig.}{Figs.}
\Crefname{figure}{Figure}{Figures}
\crefname{table}{Tab.}{Tabs.}
\Crefname{table}{Table}{Tables}
\crefname{equation}{Eq.}{Eqs.}
\Crefname{equation}{Equation}{Equations}
\crefname{appendix}{Appendix}{Appendices}
\Crefname{appendix}{Appendix}{Appendices}

\usepackage{pgfplots}

\usepackage[utf8]{inputenc}
\usepackage[T1]{fontenc}
\usepackage{mathptmx}
\usepackage{etoolbox}

\usepackage{todonotes}

\usepackage{soul}
\usepackage{xcolor}

\soulregister\cite7
\soulregister\ref7
\soulregister\eqref7
\soulregister\pageref7

\makeatletter
\def\@email#1#2{%
 \endgroup
 \patchcmd{\titleblock@produce}
  {\frontmatter@RRAPformat}
  {\frontmatter@RRAPformat{\produce@RRAP{*#1\href{mailto:#2}{#2}}}\frontmatter@RRAPformat}
  {}{}
}%
\makeatother
\begin{document}

\preprint{AIP/123-QED}

\title[Effects of anisotropic confinement on droplet rebound from superhydrophobic surfaces]{Effects of anisotropic confinement on droplet rebound from superhydrophobic surfaces}
\author{M. Feinberg}
\email{mfeinberg@ethz.ch}
\author{S.A. Hosseini}
\author{I.V. Karlin}

\affiliation{
Department of Mechanical and Process Engineering, ETH Zurich, 8092 Zurich, Switzerland
}

\homepage{https://ckg.ethz.ch.}
\date{\today}

\begin{abstract}
On flat superhydrophobic surfaces, droplet rebound is well described by a single inertio-capillary time scale, yielding a contact-time that is independent of impact energy.\cite{richardContactTimeBouncing2002,birdReducingContactTime2013}
This single-mode response reflects the radial symmetry of flat-plate impacts.
We demonstrate that an anisotropic geometric constraint, imposing a fixed spreading length along one axis, breaks this degeneracy and splits the rebound into a reciprocal pair of inertio-capillary modes.
The fixed length also couples the contact-time to the Weber-dependent maximum spread, introducing an impact-energy dependence absent on the flat plate.
We realize this constraint with grooved substrates, simulated using a non-ideal, entropic, multiple-relaxation-time lattice Boltzmann method and validated against the experiments of Chantelot \textit{et al.}\cite{chantelotWaterRingbouncingRepellent2018} 
Extending their blob model from a single transverse scale to the reciprocal pair, we organize both modes through a geometric blob number and relate their time scales to the Weber number and groove width.
We show that on non-wetting grooves the reciprocal modes are recovered directly, and explore the effects of finite wall affinity, using competition between the two modes to explain an observed two-branch structure in the contact-time response on mildly wetting, superhydrophobic grooves.
Predictions tied to global energy balance reproduce cleanly across all conditions, while those tied to the details of the droplet's spread morphology are approximate but directionally correct.
These results show that anisotropic confinement turns contact-time reduction from a question of accelerating a single rebound mode into one of selecting between conjugate inertio-capillary modes.
\end{abstract}

\maketitle

\section{\label{sec:introduction}Introduction}

The ability of superhydrophobic surfaces (SHSs) to repel and shed impacting droplets has been widely studied due to applications in industrial processes such as spray coating, self-cleaning, and anti-icing.\cite{blosseySelfcleaningSurfacesVirtual2003, krederDesignAntiicingSurfaces2016, richardContactTimeBouncing2002, lvBioinspiredStrategiesAntiicing2014, stoneIcePhobicSurfacesThat2012, wildemanSpreadingImpactingDrops2016}
Of particular interest is the duration of droplet--surface contact, which characterizes rebound dynamics and becomes especially consequential in applications such as anti-icing, where minimizing contact can reduce heat transfer, nucleation, and freezing on cold substrates.\cite{birdReducingContactTime2013,liuPancakeBouncingSuperhydrophobic2014}
These considerations have motivated extensive study of droplet-impact dynamics on SHSs, including both the physical mechanisms governing rebound and surface-design strategies for reducing contact-time.

When the wettability of the surface is sufficiently low, droplet impacts undergo a process of spreading, recoil, and takeoff analogous to the motion of a frictionless mass--spring system.\cite{richardContactTimeBouncing2002, okumuraWaterSpringModel2003} 
Consistent with this analogy, the contact-time, defined as the interval between impact and detachment, is approximately independent of the initial impact energy, commonly expressed through the Weber number $\mathrm{We} = {\rho_l U_0^2 R_0}/{\gamma}$,
which measures the ratio of inertial to surface-energy scales. 
Here $\rho_l$ is the liquid density, $\gamma$ is the surface tension coefficient, and $U_0$ and $R_0$ are the initial droplet velocity and radius, respectively.

Instead, like the oscillation period of a mass--spring system, the contact-time is set by the square root of the ratio between an effective inertia and a capillary restoring stiffness: the droplet mass and surface tension. 
Consequently $\tau_c$ scales with the quantity $\tau_0=(\rho_l R_0^3 /\gamma)^{1/2}$, known as the inertio-capillary time.\cite{richardContactTimeBouncing2002, okumuraWaterSpringModel2003}
The analogy to a single oscillator holds because the impact is radially symmetric: all in-plane spreading directions are equivalent, so the rebound is governed by one spreading–retraction coordinate, whose time scale $\tau_0$ is independent of direction.
This behavior matches the scaling law derived by Rayleigh for an oscillating drop, although the leading coefficient differs due to the asymmetries and dissipation imposed by the presence of the wall.
Experimental studies of drops impacting SHSs have found that $\tau_c \approx (2.6 \pm 0.2) \tau_0$, which differs from the Rayleigh prediction $\tau_c = \pi /\sqrt{2}\, \tau_0 \approx 2.22\, \tau_0$.\cite{richardContactTimeBouncing2002,birdReducingContactTime2013}

As in the mass--spring analogy, the oscillation amplitude remains dependent on the initial energy. 
For droplet impact, this appears in the maximum spreading radius $R_{\max}$, which increases with Weber number according to
\begin{equation}\label{eq:max_spreading_radius}
\frac{R_{\rm max}}{R_0} \propto \mathrm{We}^{1/4},
\end{equation}
in the inertial-capillary limit of negligible viscous dissipation, where $\mathrm{Re} \to \infty$ and the Ohnesorge number $\mathrm{Oh} = \sqrt{\mathrm{We}}/\mathrm{Re}$ is vanishing, $\mathrm{Oh} \to 0$.\cite{clanetMaximalDeformationImpacting2004,okumuraWaterSpringModel2003}

Numerous studies have shown that surface textures can be used to reduce the contact-time of impacting droplets.\cite{wangImpactDynamicsRebound2007, birdReducingContactTime2013, liuPancakeBouncingSuperhydrophobic2014, chantelotWaterRingbouncingRepellent2018} 
This reduction is typically achieved by reshaping the droplet during spreading and recoil. 
For example, ridges and fibers can divide the liquid into distinct lobes, each of which undergoes its own spreading and retraction process. 
The contact-time is then set by the inertio-capillary time of the largest lobe, rather than that of the full droplet, giving a reduction proportional to $(m_b/m_0)^{1/2}$, where $m_b$ and $m_0$ are the masses of the largest lobe and the original droplet, respectively.\cite{gauthierWaterImpactingSuperhydrophobic2015}

This interpretation was generalized by Chantelot \textit{et al.}\cite{chantelotWaterRingbouncingRepellent2018} through the ``blob model,'' in which the spread droplet is treated as a collection of effective sub-volumes even when distinct lobes are not visually separated.
In this framework, the mass $m_b$ of the sub-volume controlling rebound is estimated from a characteristic spreading length.
This approach has been shown to successfully describe phenomena such as ring bouncing, where a defect pierces the spreading film and causes the droplet to retract outward from a central hole, shortening the retraction distance, and allowing lift-off in a ring-like configuration.\cite{chantelotWaterRingbouncingRepellent2018}
More generally, it provides a way to connect surface-induced changes in droplet morphology to reduced contact-time.
This sub-volume interpretation has also received indirect support from recent studies of flat-surface impacts, where additional oscillation time scales have been linked to distinguishable droplet sub-volumes.\cite{liuTransitionTimeBouncing2025}

A grooved substrate provides a simple controllable way to impose an anisotropic constraint on an impacting drop: it fixes a characteristic length in one direction while leaving the drop free to spread along the other.
On this basis it was initially proposed as an experimental validation of the blob model.\cite{chantelotWaterRingbouncingRepellent2018}
Unlike the defect in the case of ring bouncing, which subdivides the spreading droplet into various ratios, the groove width remains fixed even as the maximum spreading of the droplet increases.
One consequence of this is loss of Weber independence, as has been observed in the experimental data.\cite{chantelotWaterRingbouncingRepellent2018}

A second is that the groove breaks the radial symmetry of the rebound, allowing separate spreading–retraction behaviors along the axis of the groove and normal to it. 
A fixed transverse length should therefore do more than reduce the contact-time: by lifting the directional degeneracy of the flat-plate rebound, it should split the single inertio-capillary mode into two modes with distinct time scales.
Which mode governs the observable contact-time is not necessarily fixed, but instead may be selected by the impact conditions. 
The groove geometry therefore provides a controlled setting in which to examine the general effects of anisotropy and fixed spreading lengths on droplet dynamics.

We study this using a lattice Boltzmann method for two-phase flows at high density ratio, as a fully resolved realization of the impact dynamics. 
This article is organized as follows. 
In \cref{sec:methods}, we briefly describe the lattice Boltzmann method used to simulate droplet–surface interactions, with emphasis on the wetting boundary condition, viscosity interpolation, and other model choices relevant to the present study.
In \cref{sec:setup}, we describe the simulation setup and numerical fluid properties.
\cref{sec:validation} presents a comparison of simulation results using the present method, with experimental data from Ref.~\onlinecite{chantelotWaterRingbouncingRepellent2018} for wetting and non-wetting boundaries.
In \cref{sec:groove:scaling} we present a more detailed picture of groove-induced droplet dynamics, and introduce new scaling laws developed from idealized lumped-parameter arguments that extend the blob model from a single transverse mode to a reciprocal pair of transverse and axial modes. 
We recast these modes, defined in reference to a geometric blob number, so that the relevant time scales may be inferred from the imposed Weber number and groove width.
This is followed by simulations illustrating and corroborating these scaling laws for non-wetting cases in \cref{sec:groove:nonwetting}.
Finally, in \cref{sec:groove:wetting} we investigate the effects of finite wall affinity on the same scaling phenomena, including a mode-selection transition that the two-mode framework is used to interpret.

\section{Methods}
\label{sec:methods}

The simulations were performed with an entropic, multiple-relaxation-time
(MRT) lattice Boltzmann method for two-phase flows at high density ratio,
built on a free-energy formulation. 
The scheme, its grounding in kinetic theory, and benchmarks of its thermodynamic consistency for Navier--Stokes--Korteweg dynamics are described in Refs.~\onlinecite{hosseiniConsistentLatticeBoltzmann2022,hosseiniEntropicMultirelaxationtimeLattice2022}.

The fluid is represented by discrete populations $f_i$ on a standard $D3Q27$ lattice with $D=3$ dimensions, and $Q=27$ velocities $\bm{c}_i, (i=1,\dots,Q)$ which are defined $\bm{c}_i = (c_{ix},c_{iy},c_{iz}), c_{i\alpha} \in \{-1, 0, 1\}$.
The discrete populations evolve according to
\begin{equation}
    \begin{split}
        f_i(\mathbf{r}+\mathbf{c}_i\delta t,\,t+\delta t)
        =& \left(1-\tfrac{\omega}{2}\right) f_i(\mathbf{r},t)
        + \tfrac{\omega}{2}\, f_i^{\mathrm{mirr}}(\mathbf{r},t) \\
        &+ (f_i^{\star} - f_i^{\mathrm{eq}}),    
    \end{split}
    \label{eq:lbe}
\end{equation}
where $\delta t$ is the time step size.
Here, $f_i^{\mathrm{eq}}$ is a product-form equilibrium, employed for its improved representation of third-order moments, and $f_i^{\star}$ an extended equilibrium used to introduce the non-ideal forcing.
The precise precise form of the mirror population $f_i^{\mathrm{mirr}}$ encodes the entropic MRT collision, with a central Hermite moment partitioning, and higher-order moments relaxed at a rate fixed by entropy maximization.
Full definitions of $f_i^{\mathrm{eq}}$, $f_i^{\star}$, and $f_i^{\mathrm{mirr}}$ follow
Refs.~\onlinecite{hosseiniEntropicMultirelaxationtimeLattice2022}.
The relaxation rate $\omega$ sets the kinematic viscosity $\nu$ through 
\begin{equation}
  \omega = \frac{\delta t}{\frac{\rho\nu}{P_0} + \frac{\delta t}{2}},
  \label{eq:omega}
\end{equation}
with $P_0 = \rho\varsigma^2$ the ideal-gas reference pressure carried in the equilibrium and $\varsigma = \delta r/\sqrt{3}\,\delta t$ the lattice speed of sound.

As only the ideal-gas pressure is retained in the equilibrium, non-ideal thermodynamics and capillarity enter through a body force $\bm{F}$ applied at the population level as the source term $(f_i^\star - f_i^{\mathrm{eq}})$ according to the exact-difference method of Kupershtokh.\cite{kupershtokhNewMethodIncorporating2004}
Non-ideal thermodynamics are modeled using the van der Waals equation of state
\begin{equation}
  P = \frac{\rho R T}{1-b\rho} - a\rho^2,
  \qquad
  a = \frac{27}{64}\frac{R^2 T_c^2}{P_c},
  \quad
  b = \frac{1}{8}\frac{R T_c}{P_c},
  \label{eq:vdw}
\end{equation}
where $a$ and $b$ are fixed by the critical state $(\rho_c, T_c, P_c)$.
The non-ideal contribution is introduced as a force $\nabla(P-P_0)$, which is recast through a pseudo-potential $\psi$ using the identity $\nabla \phi = 2 \sqrt{\phi} \nabla \sqrt{\phi}$:
\begin{equation}
  \psi =
  \begin{cases}
    \sqrt{P-P_0}, & P > P_0,\\[2pt]
    \sqrt{P_0-P}, & P \le P_0,
  \end{cases}
  \label{eq:psi}
\end{equation}
which reduces gradient magnitudes and hence the discretization error in the
pressure-gradient evaluation.
Combining this with the diffuse-interface Korteweg contribution gives the body force
\begin{equation}
  \mathbf{F} =
  \begin{cases}
    \phantom{-}2\psi\nabla\psi - \kappa\rho\nabla\nabla^2\rho, & P > P_0,\\[2pt]
    -2\psi\nabla\psi - \kappa\rho\nabla\nabla^2\rho, & P \le P_0,
  \end{cases}
  \label{eq:korteweg}
\end{equation}
where the capillarity coefficient $\kappa$ provides independent control of
the surface tension.
All spatial derivatives are evaluated by finite differences.

To mirror the large viscosity contrast between phases in experimental droplet--surface interactions, a phase-dependent viscosity is used.
Here, and throughout, subscripts $l$ and $v$ denote the liquid and vapor phases, so that $\rho_l$, $\rho_v$, $\nu_l$ and $\nu_v$ are the phase-specific densities and kinematic viscosities.
A phase indicator $\alpha_{\rm visc}(\rho) = (\rho-\rho_v)/(\rho_l-\rho_v)$, linear in the local density, is defined and clamped to $\alpha_{\rm visc}\in[0,1]$.
The local viscosity is interpolated harmonically,
\begin{equation}
  \frac{1}{\nu(\alpha_{\rm visc})}
  = \left[
      \left(\frac{\alpha_{\rm visc}}{\nu_l}\right)^2
    + \left(\frac{1-\alpha_{\rm visc}}{\nu_v}\right)^2
    \right]^{1/2},
  \label{eq:visc}
\end{equation}
with a fixed ratio $\nu_v/\nu_l = 15$ throughout.

Wetting is imposed by assigning a fictitious density $\rho_w \in [\rho_v, \rho_l]$ to solid nodes.\cite{hosseiniConsistentLatticeBoltzmann2022,benziMesoscopicModelingTwophase2006,sbragagliaSurfaceRoughnessHydrophobicityCoupling2006}
Through the non-local evaluation of the Korteweg force, this wall density modifies the capillary forcing in adjacent fluid nodes and thereby sets the wall affinity.
The lower limit $\rho_w=\rho_v$ gives a fully non-wetting wall ($\theta_c=180^\circ$), while increasing $\rho_w$ increases the affinity, with the limit $\rho_w \to \rho_l$ corresponding to the hydrophilic limit in this parameterization.
Wetting conditions are reported below by their measured static contact angle $\theta_c$. 
The no-slip/no-penetration condition at the solid boundary is enforced using a standard halfway bounce-back boundary treatment.\cite{krugerLatticeBoltzmannMethod2017}

\section{\label{sec:setup}Simulation Setup}

A series of simulations were performed for droplets impacting flat plates and grooved substrates of varying groove width.
In all cases the computational domain was a rectangular prism with the origin at the nominal point of impact.
The droplet was initialized with velocity $\bm{u}=(0,0,-U_0)$, so that impact occurred along the negative $z$-direction onto a solid substrate at $z=0$.
A zero-gradient condition was imposed at the upper boundary and periodic conditions in the remaining directions.
For grooved substrates, the groove axis was aligned with the $y$-direction, the groove width is denoted by $W$, and the groove height was fixed at $H/R_0=4/3$.
The precise value of $H$ is not essential, provided it is sufficient to properly constrain transverse spreading.

\begin{figure}
    \includegraphics[width=\linewidth]{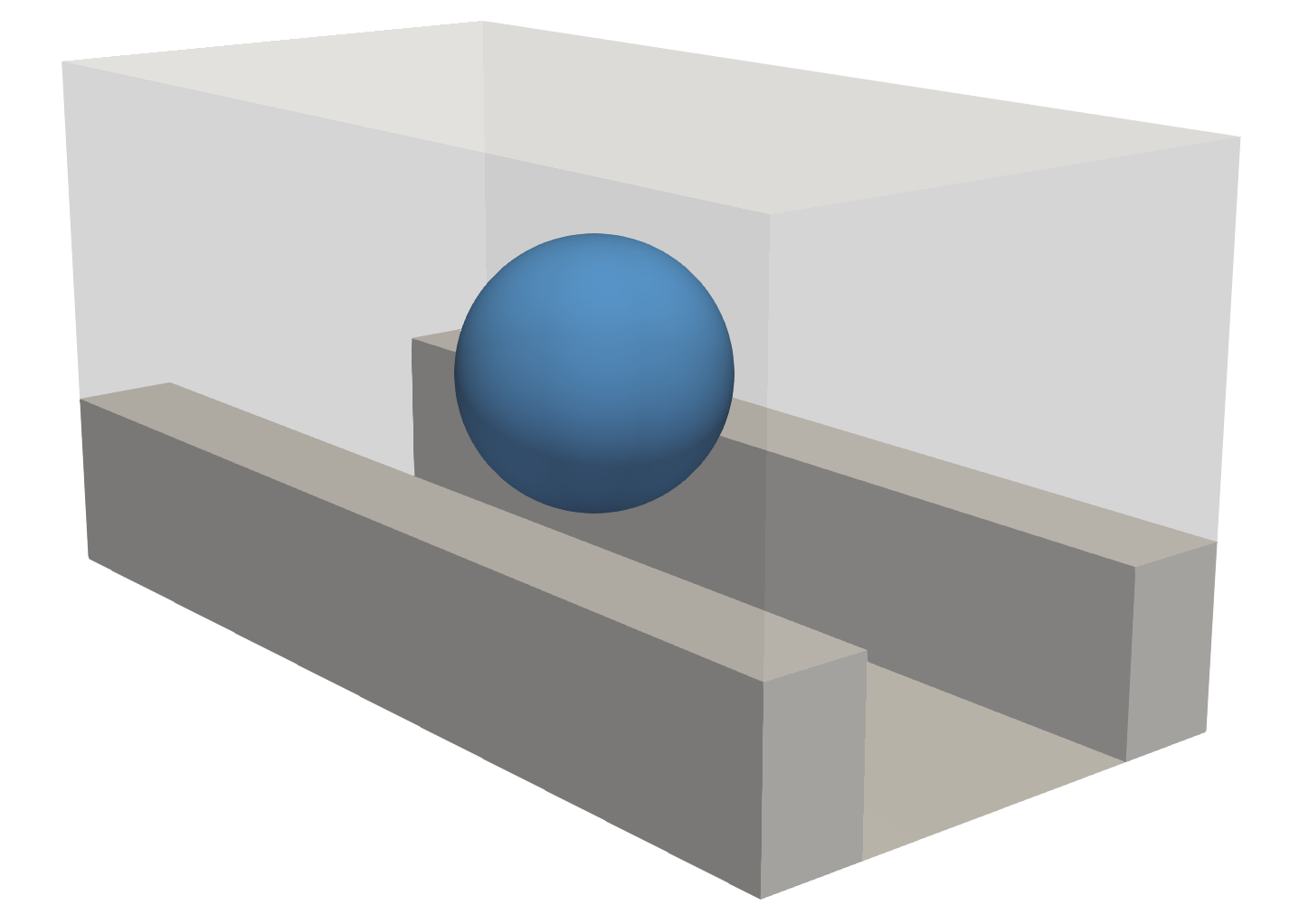}
    \caption{Simulation domain for a grooved substrate with
    $W/R_0=2.5$.}
    \label{fig:domain}
\end{figure}

Domain sizes were parameterized by the initial droplet radius $R_0$ and chosen to prevent droplet--boundary interaction over the course of the collision.
The base grooved-substrate domain was $4R_0 \times 8R_0 \times 4R_0$, as illustrated in \cref{fig:domain}, with the longer dimension along the groove axis.
Flat-plate impacts used $8R_0 \times 8R_0 \times 6R_0$, and selected
runs for assessing post-detachment oscillations used an enlarged $6R_0 \times 10R_0 \times 10R_0$ domain to accommodate axial elongation and several unconstrained
oscillation periods.
All simulations used $R_0 = 64\,\delta r$, and were typically terminated once the droplet had fully detached from the substrate.
Contact was identified by tracking the lowest point at which the density exceeded the mean liquid–vapor density and comparing its distance from the wall with a threshold based on the diffuse-interface width.
For cases involving temporary detachment and reattachment, particularly under non-wetting conditions, the contact-time was defined as the interval between first and final contact.

The fluid properties were set in non-dimensional lattice units, tuned to provide the desired density ratio and interface representation.
A density ratio of $\rho_l/\rho_v \approx 600$ was targeted for agreement with the
experiments of Ref.~\onlinecite{chantelotWaterRingbouncingRepellent2018}.
From Maxwell's construction, the corresponding reduced temperature was found to be $T/T_c = 0.3936$. 
The van der Waals parameters $a$ and $b$ were then selected to give a converged interface of minimal width consistent with stability, accurate coexistence densities, and low spurious currents. 
The resulting constants are listed in \cref{tab:fluid_constants}.

The gradients entering the Korteweg force in \cref{eq:korteweg} are poorly resolved when the diffuse interface spans only a few grid nodes.
In this regime, third-order errors in the pressure-gradient discretization produce an additional numerical contribution to the surface tension.
To offset this excess numerical surface tension, we employ a negative capillary coefficient, $\kappa=-0.0025$, which lowers the measured surface tension and permits a thinner stable interface representation.
The thinner interface increases the accessible Weber number by delaying artificial breakup in highly stretched lamellae, where opposing diffuse interfaces can overlap once the film thickness becomes comparable to the interface breadth.
The resulting surface tension coefficient is measured from the Laplace pressure of static liquid columns, giving $\gamma \approx 0.0704$.
Details of these measurements and validation of the surface tension behavior of the model are presented in \cref{app:surfacetension}.

\begin{table}
    \centering
    \begin{tabular}{c c c}
    \hline
    Parameter & Symbol & Value \\
    \hline
    Specific gas constant & $R$ & 1.0 \\
    Attraction parameter & $a$ & 0.00908 \\
    Excluded-volume parameter & $b$ & 0.0952 \\
    Capillarity coefficient & $\kappa$ & $-0.0025$ \\
    Kinematic viscosity (liquid) & $\nu_l$ & 0.025 \\
    Kinematic viscosity (vapor) & $\nu_v$ & 0.375 \\
    Reduced temperature & $T/T_c$ & 0.3936 \\
    Density (liquid) & $\rho_l$ & 9.088 \\
    Density (vapor) & $\rho_v$ & 0.015 \\
    Density (wall) & $\rho_w$ & \{0, 0.695\} \\
    Surface tension & $\gamma$ & 0.0704 \\
    \hline
    \end{tabular}
    \caption{\label{tab:fluid_constants}Fluid constants used in the
    simulations. 
    All values are in non-dimensional lattice units.
    The wall density $\rho_w$ shows the two values used for wetting and non-wetting boundary treatments. 
    $\gamma$ is a measured, not imposed, quantity as described in \cref{app:surfacetension}.}
\end{table}

Two solid boundary conditions were employed.
The first is fully non-wetting ($\rho_w = \rho_v$, $\theta_c \approx 180^\circ$), used as an approximation of the heated, Leidenfrost-like boundary in the experiments of Ref.~\onlinecite{chantelotWaterRingbouncingRepellent2018}.
The second is a partially wetting superhydrophobic condition ($\rho_w = 0.695$, $\theta_c = 165^\circ \pm 3^\circ$). 
These static contact angles were measured from simulations of a liquid column confined between two parallel solid walls, as described in \cref{app:wetting}.

For each geometry, simulations were run over a range of Weber numbers by varying the impact velocity $U_0$ while holding $\nu_l$, $\gamma$, and $R_0$ fixed.
The Ohnesorge number is fixed at $\mathrm{Oh}=\sqrt{\mathrm{We}}/\mathrm{Re}=0.036$ across all series, so that variations reflect changes in impact inertia alone. 
The simulated value is greater than is typical for millimetric water drops, but within the inertio-capillary regime, as confirmed by the recovery of the canonical $\tau_c \approx 2.6 \tau_0$ on the flat plate in \cref{sec:validation}.

\section{\label{sec:validation}Comparison with experimental results}

The experiments of Ref.~\onlinecite{chantelotWaterRingbouncingRepellent2018} employed two repellent surfaces.
The first was a silicon wafer coated with hydrophobic nanobeads, giving a static contact angle of $166^\circ \pm 4^\circ$; we refer to this, hereafter, as the superhydrophobic surface (SHS).
The second was a brass plate heated to $350^\circ \mathrm{C}$, hot enough to sustain a Leidenfrost vapor layer beneath the impacting drop; we refer to this as the Leidenfrost surface.
These cases are represented in the simulations by the two boundary treatments introduced in \cref{sec:setup}.
The SHS is modeled using the partially wetting boundary condition, with measured contact angle $\theta_c \approx 165^\circ$, in close agreement with experiment. 
The Leidenfrost surface is modeled using the fully non-wetting condition, $\rho_w=\rho_v$ and $\theta_c\approx180^\circ$, which approximates the geometric separation produced by the vapor layer but does not include thermal effects.

\begin{figure*}[t]
    \centering
    \includegraphics{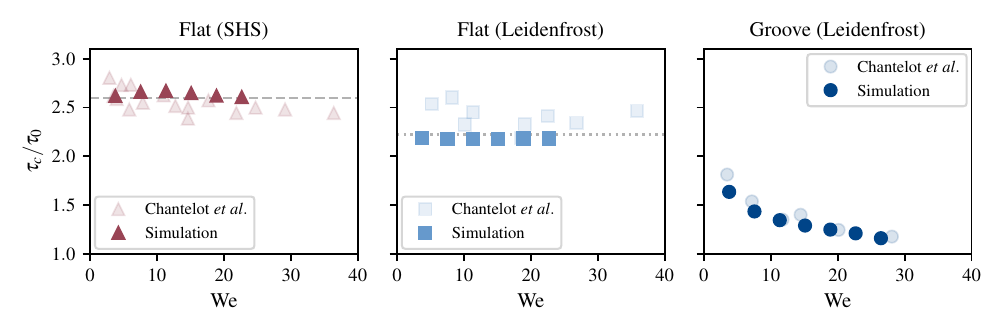}
    \caption{Comparison of simulation and experimental contact-times from Ref.~\onlinecite{chantelotWaterRingbouncingRepellent2018} across three configurations. 
    Faded markers are experiments, and saturated markers simulations. 
    \textbf{Left}: droplets impacting a flat superhydrophobic surface (nanobead-coated silicon wafer in experiment, $\theta_c \approx 165^\circ$ in simulation). 
    The dashed line marks the literature value $\tau_c = 2.6\, \tau_0$.\cite{chantelotWaterRingbouncingRepellent2018,birdReducingContactTime2013, richardContactTimeBouncing2002} 
    \textbf{Center}: droplets impacting a flat Leidenfrost surface (brass at $350^\circ \mathrm{C}$ in experiment, $\theta_c = 180^\circ$ in simulation).
    The dotted line marks the Rayleigh free-oscillation limit $(\pi/\sqrt{2})\, \tau_0 \approx 2.22\, \tau_0$. 
    \textbf{Right}: droplets impacting a Leidenfrost groove of width $W/R_0 = 2.37$ (brass at $350^\circ \mathrm{C}$ in experiment, $\theta_c = 180^\circ$ in simulation).}
    \label{fig:exp-v-sim-tc}
\end{figure*}

\begin{figure*}
    \centering
    \includegraphics[width=\textwidth]{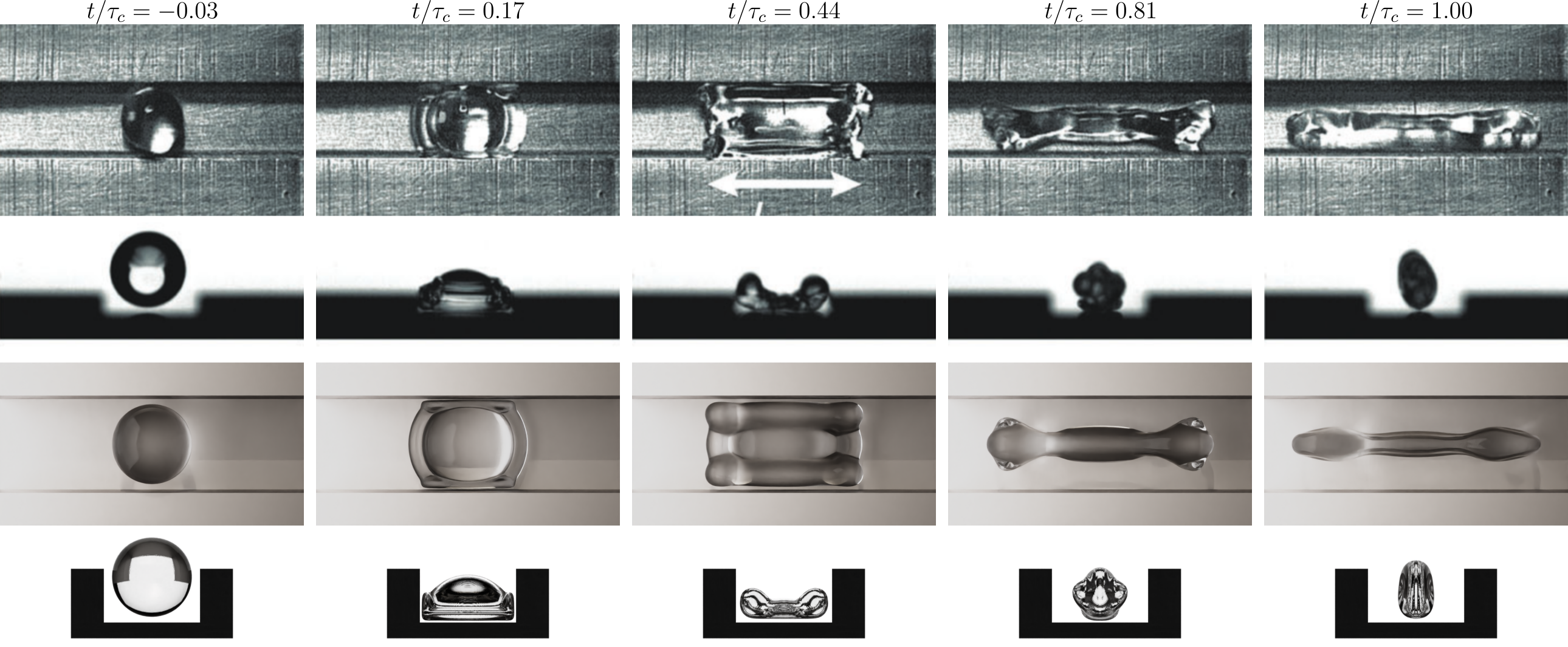}
    \caption{Droplet impact at $\mathrm{We} \approx 20.1$ in a groove of $W/R_0 = 2.37$; \textbf{(top)} Experimental images reproduced from Ref.~\onlinecite{chantelotWaterRingbouncingRepellent2018} with permission from the Royal Society of Chemistry; \textbf{(bottom)} corresponding simulation results at the nearest output time.}
    \label{fig:exp-v-sim-images}   
\end{figure*}

To assess the suitability of the numerical scheme for the present application, contact-times were measured over a range of Weber numbers for three configurations reported in Ref.~\onlinecite{chantelotWaterRingbouncingRepellent2018}: a flat SHS, a flat Leidenfrost surface, and a Leidenfrost groove of width $W/R_0=2.37$. 
No wetting-groove case was reported experimentally.
Chantelot \textit{et al.} state that ``water impacting [an SHS] groove pins along the edges, which artificially modifies and/or scatters the contact-time,'' and therefore restricted their grooved-substrate measurements to the Leidenfrost surface.\cite{chantelotWaterRingbouncingRepellent2018} 
Contact-time results are presented in \cref{fig:exp-v-sim-tc}, and a qualitative comparison of the droplet evolution for the groove case is shown in \cref{fig:exp-v-sim-images}.

For the flat SHS surface, the simulated contact-time averages $2.63\, \tau_0$ across $\mathrm{We} \in [3.7, 22.7]$ with no systematic dependence on the Weber number. 
This is in close agreement with the experimental average of $2.56\, \tau_0$ and consistent with the value $(2.6 \pm 0.2) \tau_0$ accepted in the literature.\cite{richardContactTimeBouncing2002, birdReducingContactTime2013, chantelotWaterRingbouncingRepellent2018}
The recovery of both the Weber-independent scaling and the correct pre-factor confirms that the wetting boundary condition reproduces the canonical inertio-capillary behavior expected on a superhydrophobic surface.

For the flat Leidenfrost surface, the experiments in Ref.~\onlinecite{chantelotWaterRingbouncingRepellent2018} show a reduced, Weber-independent contact-time, with mean value $\tau_c \approx 2.41\,\tau_0$. 
The simulations reproduce the Weber invariance but predict a stronger reduction, giving $\tau_c \approx 2.19\,\tau_0$, close to Rayleigh's inviscid free-oscillation time $\pi/\sqrt{2}\,\tau_0 \approx 2.22\,\tau_0$. 
Thus, the simulations capture the qualitative Leidenfrost-like signature of reduced, Weber-independent contact-time, but moderately under-predict the wall influence observed experimentally.

The most likely source of this discrepancy is that the Leidenfrost effect is not merely a passive non-wetting boundary condition, but a thermally sustained vapor layer with coupled heat, mass, pressure, and flow fields.
The present isothermal boundary treatment mimics the geometric separation associated with a vapor layer, but cannot capture these coupled dynamics.
One alternative source of discrepancy was tested directly: the experimental Bond number $\mathrm{Bo} = \Delta\rho g R_0^2/\gamma$, which describes the ratio of gravitational to capillary forces, is small but non-negligible at $\mathrm{Bo} \approx 0.354$. 
If gravity influenced the drop's adherence to the surface, it could in principle prolong the contact-time independent of any inertial effect. 
A simulation repeated with a corresponding gravitational force imposed produced no meaningful change in contact-time, ruling out gravity as the significant source of the offset.

For the groove geometry, simulations recover quantitative agreement with experiment, closely matching both the magnitude and trend of $\tau_c$ across $\mathrm{We} \in [3.7,26.4]$. 
The same $\theta_c \approx 180^\circ$ boundary that under-predicts contact-time on a flat Leidenfrost surface therefore produces improved accuracy in the groove configuration, with the largest discrepancies confined to low Weber numbers.

The implication is that contact-time in the two geometries is controlled by different physics. 
On a flat surface, deviations from Rayleigh-like free oscillation arise from the sustained tangential interaction between the spreading liquid and the substrate, so the measured contact-time remains sensitive to wetting, slip, friction, and vapor-film dynamics. 
In the groove, by contrast, the dominant time scale is set by the inertial redirection of liquid by the groove walls and the subsequent transverse recoil. 
It is therefore less sensitive to the detailed droplet--substrate contact physics. 
Accordingly, the non-wetting boundary is sufficient for the regime considered here.
Wetting effects in the groove geometry are considered in more detail in \cref{sec:groove:wetting}.

\section{\label{sec:groove}Groove--droplet dynamics}

\begin{figure}
    \centering
    \includegraphics[width=\linewidth]{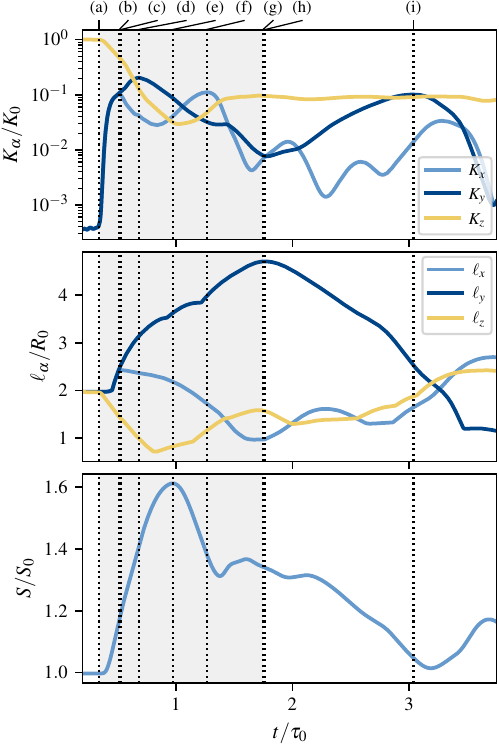}
    \caption{Evolution of a droplet with $\mathrm{We}=11.4$ interacting with a groove of width $W/R_0=2.5$.
    Directions subscripted as $\alpha \in \{x,y,z\}$ refer to the transverse, axial, and vertical directions respectively.
    The gray shaded region corresponds to the time interval in which the droplet is in contact with the groove.
        \textbf{(top)} Directional kinetic energies $K_\alpha/K_0$, 
        \textbf{(middle)} droplet extents $\ell_\alpha/R_0 = (\alpha_{\max} - \alpha_{\min})/R_0$, and 
        \textbf{(bottom)} droplet surface area.
    Marked events are: 
        \textbf{(a)} first contact with the groove; 
        \textbf{(b)} first peak in $K_x$, corresponding to the strongest transverse spreading motion; 
        \textbf{(c)} first contact with the groove walls; 
        \textbf{(d)} first peak in $K_y$, corresponding to the strongest axial spreading motion; 
        \textbf{(e)} maximum surface area $S_{\max}$, where $\ell_{\max}$ is measured; 
        \textbf{(f)} transverse rebound, identified by the second peak in $K_x$; 
        \textbf{(g)} droplet release;
        \textbf{(h)} peak axial length of the droplet;
        \textbf{(i)} axial recoil, identified by the second peak in $K_y$.
    }
    \label{fig:droplet-evolution}
\end{figure}

To characterize the droplet--groove interaction, we track the droplet extents, surface area, and directional kinetic energies,
\begin{equation}
    K_\alpha = \frac{1}{2}\int_V \rho_l u_\alpha^2\,dV ,
\end{equation}
where $V$ is the droplet volume, $\alpha \in \{x,y,z\}$, and $\sum_\alpha K_\alpha$ is the total droplet kinetic energy. 
Representative time histories for a droplet at $\mathrm{We}=11.3$ impacting a $W/R_0=2.5$ groove are shown in \cref{fig:droplet-evolution}, with letters marking key moments.

The interaction with the groove, the shaded region in \cref{fig:droplet-evolution}, proceeds through three stages. 
First, the droplet undergoes axisymmetric spreading from initial contact at (a) until the liquid reaches the groove walls at (c). 
The duration of this stage depends on both Weber number and groove width: higher-$\mathrm{We}$ impacts spread faster, while wider grooves require farther spreading before wall contact. 
Because wall contact interrupts the outward motion before a full inertio-capillary oscillation is completed, the wall-contact-time is set by the spreading velocity rather than by an amplitude-independent oscillation period. 
In the flat-plate limit, $W/R_0 \to \infty$, or low energy limit $\mathrm{We} \to 0$, the walls are never reached and Weber-independent rebound is recovered.

The second stage is characterized by groove-induced anisotropic spreading.
Once the transverse motion is arrested, kinetic energy is redirected from the transverse to the axial direction. 
This stage extends approximately from the first peak in $K_x$ at (b), which marks the strongest transverse spreading motion, to the second peak in $K_x$ at (f), which marks the moment of peak transverse recoil. 
Within this interval, the droplet reaches its maximum surface area $S_{\max}$ at (e), and its maximum axial spreading motion, as measured by a peak in $K_y$ at (d).

The third stage is the departure stage, beginning with rapid transverse recoil at (f) and ending with lift-off from the groove at (g).
The droplet reaches its maximum axial extent at (h).
This is not in general coincident with (g), and depends on Weber number, however, the droplet typically departs the groove in an axially extended configuration.
Finally, a second peak in $K_y$ occurs at (i) corresponding to rapid axial recoil, which occurs after the droplet has departed the groove.

\subsection{\label{sec:groove:scaling}Droplet-groove scaling laws}

The central observation of the blob model of Chantelot \textit{et al.}\cite{chantelotWaterRingbouncingRepellent2018} is that, for a droplet rebounding from a groove, the transverse spreading and retraction of the droplet determines the contact-time. 
The model therefore decomposes the droplet so that the largest effective blob is defined as having a maximum spreading length equal to the groove-imposed transverse length scale $W$. 
At maximum spread, the droplet is approximated as a rectangular prism of width $W$, axial length $\ell_{\max}$, and thickness $h$, with mass conservation requiring $W\,\ell_{\max}\,h \sim R_0^3$.
Here $\ell_{\max}$ is the axial length of the droplet at the moment of maximum surface area, marked (e) in \cref{fig:droplet-evolution}, not the maximum axial extension (h).

This prismatic volume is decomposed into $N=\ell_{\max}/W$ roughly isotropic blobs distributed along the groove axis. 
Each blob has mass $m_b=m_0/N$, where $m_0$ is the initial droplet mass. 
Applying inertio-capillary scaling to the effective blob gives
\begin{equation}\label{eq:tc-scaling}
\frac{\tau_\perp}{\tau_0}
\sim
\left(\frac{m_b}{m_0}\right)^{1/2}
=
\left(\frac{W}{\ell_{\max}}\right)^{1/2}
=
N^{-1/2},
\end{equation}
where $\tau_\perp$ is the time scale associated with transverse spreading and retraction. 
In the original blob-model interpretation, this transverse time scale controls the contact-time, $\tau_c \sim \tau_\perp$. 
This scaling has been observed in experimental data for a single groove width.\cite{chantelotWaterRingbouncingRepellent2018}

However, the transverse mode $\tau_\perp$ is not the only groove-imposed time scale.
As can be seen in the kinetic-energy decomposition in \cref{fig:droplet-evolution}, the spreading and retraction dynamics in the directions perpendicular and parallel to the groove are decoupled. 
The transverse mode controls retraction from the groove walls, while a second, longer time scale governs axial recoil of the elongated droplet. 
We denote this axial mode by $\tau_\parallel$. 

A scaling for $\tau_\parallel$ may be derived by analogy with Taylor--Culick retraction.\cite{taylorDynamicsThinSheets1959,culickCommentsRupturedSoap1960}
For a film of thickness $h$, the free edge retracts at the capillary-inertial velocity
\begin{equation}
V_{TC} = \sqrt{\frac{2\gamma}{\rho_l h}}.
\end{equation}
Taking the relevant axial length scale to be $\ell_{\max}$ gives
\begin{equation}
    \tau_\parallel \sim \frac{\ell_{\max}}{V_{TC}} .
\end{equation}
Substituting $h \sim R_0^3/(W\ell_{\max})$ and dropping factors of order unity yields
\begin{equation}
    \tau_\parallel
    \sim
    \left(
    \frac{\rho_l R_0^3}{\gamma}
    \frac{\ell_{\max}}{W}
    \right)^{1/2}
    =
    \tau_0
    \left(
    \frac{\ell_{\max}}{W}
    \right)^{1/2},
\end{equation}
and consequently
\begin{equation}\label{eq:ta-scaling}
    \frac{\tau_\parallel}{\tau_0} \sim N^{1/2}.
\end{equation}
The groove therefore imposes a pair of inversely scaling inertio-capillary modes,
\begin{equation}\label{eq:mode-product}
    \tau_\perp \tau_\parallel \sim \tau_0^2,
\end{equation}
which replace the single degenerate rebound time scale of the flat-plate interaction. 

Although originally defined for free liquid sheets, the Taylor--Culick relation has been applied previously to the recoil of spread films and lamellae in droplet-impact problems.\cite{birdReducingContactTime2013,josserandDropImpactSolid2016,wildemanSpreadingImpactingDrops2016} 
It follows from a thin-film force balance and is best justified at higher Weber numbers, where the droplet most convincingly approximates an extended sheet between the groove walls. 
An alternative derivation, presented in Appendix~\ref{ap:axial-mode-inertial}, based on capillary-driven inertial recoil gives the same scaling up to $\mathcal{O}(1)$ constants, without relying on the specific Taylor--Culick rim-retraction mechanism. 
The agreement of these two arguments adds further support for \cref{eq:ta-scaling}.

To express the blob number in terms of imposed parameters, the measured length $\ell_{\max}$ may be related to $W$ and $\mathrm{We}$. 
We assume that, to leading order, the groove redistributes the spreading anisotropically, while the overall conversion of impact kinetic energy into capillary energy, which sets the maximum spread, is left unchanged.
The flat-plate spreading law in \cref{eq:max_spreading_radius} may therefore be used to estimate the spread surface area.
A droplet impacting a flat plate forms, at maximum spread, a geometry analogous to a flattened disk of radius $R_{\max}$.
The surface area is therefore dominated by the two large faces, which implies $S_{\max}\sim R_{\max}^2$.
Combined with the radius law $R_{\max}/R_0\sim \mathrm{We}^{1/4}$ of \cref{eq:max_spreading_radius} we can write
\begin{equation}\label{eq:smax-scaling}
    \frac{S_{\max}}{R_0^2}
    \sim
    \left(\frac{R_{\max}}{R_0}\right)^2
    \sim
    \mathrm{We}^{1/2}.
\end{equation}
We note that this scaling, derived from flat-plate arguments, carries no groove-width dependence. 
Adopting it for the groove geometry therefore assumes that confinement reshapes the spread without affecting the global conversion of kinetic to capillary energy that sets $S_{\max}$.
This is a strong assumption --- narrower grooves impose more rapid transverse arrest, and there is no fundamental reason to expect this to leave the total spread area unaffected. 
We adopt the assumption to make progress, and return to it in \cref{sec:groove:discussion} when interpreting the resulting scaling laws against the data.

For the rectangular blob geometry, the same thin-film approximation, valid while $h \ll W, \ell_{max}$, gives 
\begin{equation}\label{eq:rectangular-area-assumption}
    S_{\max} \sim W\,\ell_{\max},
\end{equation}
and consequently
\begin{equation}\label{eq:length-We-scaling}
    \frac{\ell_{\max}}{R_0}
    \sim
    \frac{R_0}{W}\mathrm{We}^{1/2}.
\end{equation}
The blob number can therefore be expressed in terms of the imposed parameters as
\begin{equation}\label{eq:N-scaling}
N \sim \frac{\mathrm{We}^{1/2}}{(W/R_0)^2}.
\end{equation}
Combining \cref{eq:N-scaling} with \cref{eq:tc-scaling,eq:ta-scaling} gives the twin scaling predictions
\begin{equation}\label{eq:tc-We-scaling}
    \frac{\tau_\perp}{\tau_0}
    \propto
    \frac{W/R_0}{\mathrm{We}^{1/4}},
\end{equation}

and
\begin{equation}\label{eq:ta-We-scaling}
    \frac{\tau_\parallel}{\tau_0}
    \propto
    \frac{\mathrm{We}^{1/4}}{W/R_0}.
\end{equation}
Comparison with \cref{eq:max_spreading_radius} implies a geometric interpretation of \cref{eq:tc-We-scaling,eq:ta-We-scaling}. 
The transverse and axial time scales are controlled by the ratio between the groove width $W$ and the natural flat-plate spreading radius $R_{\max}$, up to constants and the approximations used in the geometric closure.
Thus the two modes vary inversely with the extent to which the groove impinges upon the droplet’s otherwise free spreading.

Although the blob-model scalings should lose quantitative accuracy as $N\to1$, the approach to the flat-plate limit is expected to be continuous, with $\tau_\perp$ and $\tau_\parallel$ merging as groove-imposed effects vanish. 
This suggests comparable prefactors for the two modal scalings, so that $\tau_\perp < \tau_\parallel$ for $N>1$ --- an ordering confirmed by simulations in \cref{sec:groove:nonwetting}. 
Because lift-off is caused by in-plane recoil redirecting into vertical momentum, the idealized inertio-capillary picture predicts that the faster transverse retraction governs lift-off, giving $\tau_c \sim \tau_\perp \propto W\,\mathrm{We}^{-1/4}$.

\subsection{\label{sec:groove:nonwetting}Droplet impacts on non-wetting grooves}

To test the predictions of \cref{eq:tc-scaling,eq:tc-We-scaling,eq:ta-scaling}, simulations were run for grooves with dimensions
$W/R_0 \in \{2.25, 2.37, 2.5, 2.75, 3.0\}$ across $\mathrm{We} \in [3,30]$ using the non-wetting boundary condition. 
Simulations produced consistent qualitative trends: contact-time decreases monotonically with both Weber number and groove width. 
The widest groove at its lowest Weber number, $W/R_0=3.0$ and $\mathrm{We}=3.8$, saturates to the flat-plate contact-time, with the droplet only barely reaching the groove walls.

\begin{figure}
    \centering
    \includegraphics{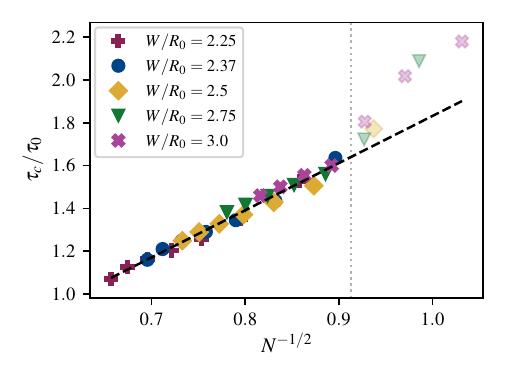}
    \caption{Groove contact-times as a function of $N^{-1/2}$, where $N=\ell_{\max}/W$. 
    Data from all five groove widths collapse onto a common linear trend at sufficiently large $N$, but depart from this scaling as $N\to1$, where $\ell_{\max}$ becomes comparable to $W$ and the groove walls no longer strongly constrain the dynamics. 
    Faded data points are those excluded from the least squares fit (dashed black line), and the dotted line marks the threshold $N=1.2$, below which data is excluded here, and in subsequent fits.
    }
    \label{fig:tc-scaling}
\end{figure}

We begin by testing \cref{eq:tc-scaling}, since it follows directly from the blob decomposition. 
Using $N=\ell_{\max}/W$, with $\ell_{\max}$ measured at the moment of maximum surface area, we plot $\tau_c/\tau_0$ against $N^{-1/2}$ in \cref{fig:tc-scaling}. 
Data from all five groove widths collapse onto a linear trend in $N^{-1/2}$, confirming the scaling predicted by \cref{eq:tc-scaling} when the droplet is sufficiently constrained by the groove.
The collapse begins to deteriorate only as $N$ approaches unity, with a visible departure near $N\approx1.2$, or equivalently $N^{-1/2}\approx0.91$. 
Below this threshold, the data depart toward the flat-plate contact-time.
This deviation is expected: when $\ell_{\max}$ is only slightly larger than $W$, the spread droplet cannot be reliably partitioned into multiple roughly isotropic blobs, and the rectangular-prism geometry assumed in deriving \cref{eq:tc-scaling,eq:ta-scaling} is not realized.
The droplet therefore behaves increasingly like an unconstrained flat-plate impact. 
We use this observed departure from linearity, rather than an a priori cutoff, to define the in-regime data set as $N>1.2$.

\begin{figure}
    \centering
    \includegraphics{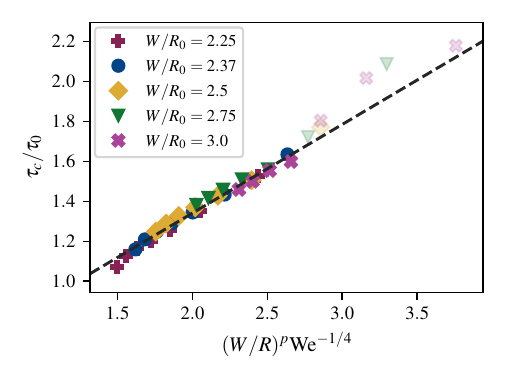}
    \caption{contact-time $\tau_c/\tau_0$ versus the scaling variable $(W/R_0)^p\, \mathrm{We}^{-1/4}$ for all five groove widths, plotted on a single axis to assess the collapse predicted by \cref{eq:ls-fit-form}. The solid line shows the fit obtained from in-regime data ($N > 1.20$), with $C = 0.44$, $b = 0.45$, $p = 1.51$. Faded markers indicate excluded data, in which the droplet behavior is transitioning towards the flat-plate limit.}
    \label{fig:tc-we-scaling-collapse}
\end{figure}

\begin{figure}
    \centering
    \includegraphics{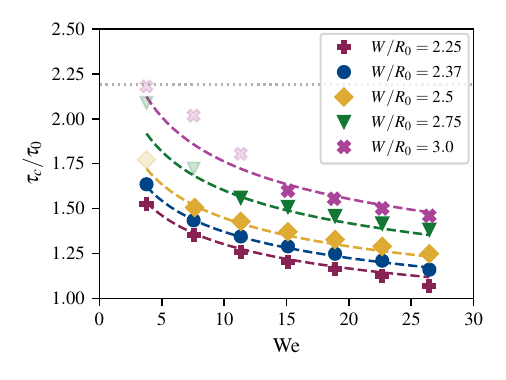}
    \caption{contact-time $\tau_c/\tau_0$ as a function of Weber number for each groove width, with dashed curves showing the per-width prediction of the fitted correlation \cref{eq:ls-fit-form}. 
    The $\mathrm{We}^{-1/4}$ dependence is well reproduced across all groove widths within the regime of validity of the fit. Faded markers indicate data with $N < 1.20$, excluded from the fit, in which the droplet behavior is transitioning towards the flat-plate limit (marked with a dotted grey line).}
    \label{fig:tc-we-scaling}
\end{figure}

Having confirmed the measured-$N$ scaling of \cref{eq:tc-scaling}, we next test the imposed-parameter form of \cref{eq:tc-We-scaling}, which expresses the contact-time in terms of $W/R_0$ and $\mathrm{We}$ alone.
We fit the data using
\begin{equation}\label{eq:ls-fit-form}
    \frac{\tau_c}{\tau_0}
    =
    b + C\, (W/R_0)^p\, \mathrm{We}^{-1/4}.
\end{equation}
The fit is performed on the in-regime data ($N>1.2$) by unweighted least squares.
At fixed $p$, the model is linear in $(C,b)$, so the inner fit reduces to a standard linear regression on the scaling variable $(W/R_0)^p\,\mathrm{We}^{-1/4}$. 
The exponent $p$ is then chosen to minimize the residual sum of squares using a bounded Brent optimization method.
We obtain $C=0.44$, $b=0.45$, and $p=1.51$, with $\mathrm{RMSE}=0.02$ and $R^2=0.98$. 
Collapsed data and fit are shown in \cref{fig:tc-we-scaling-collapse}.
Per-width contact-time data and curves are shown in \cref{fig:tc-we-scaling}.

The $\mathrm{We}^{-1/4}$ dependence is well reproduced across all in-regime groove widths.
Groove width dependence is monotonic and systematic; however, the fitted exponent is significantly larger than the linear dependence predicted by \cref{eq:tc-We-scaling}. 
Repeating the fit with each groove width removed in turn gives $p$ between $1.45$ and $1.68$, with the four narrower omissions clustered between $1.45$ and $1.53$ and the shift to $1.68$ corresponding to removal of the widest groove. 
The data therefore support the qualitative conclusion that the effective groove-width dependence is super-linear, while placing only loose bounds on the precise exponent.

\begin{figure}
    \centering
    \includegraphics{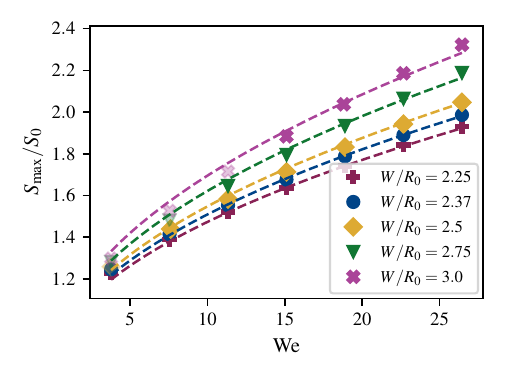}
    \caption{Maximum droplet surface area as a function of $\mathrm{We}$ for each groove width. 
    Dashed curves show least-squares fits of \cref{eq:ls-fit-sa-form}, giving $C=0.11$, $b=0.76$, $p=0.94$, $\mathrm{RMSE}=0.01$, and $R^2=0.99$.
    The data retain the predicted $\mathrm{We}^{1/2}$ dependence, but with an additional groove-width dependence.
    Faded markers indicate data with $N < 1.20$, excluded from the fit.}
    \label{fig:smax-scaling}
\end{figure}

To identify the source of the super-linear groove-width dependency, we examine the geometric scalings used to eliminate $N$, namely those for $S_{\max}$ and $\ell_{\max}$. 
The maximum surface area, shown in \cref{fig:smax-scaling}, retains the predicted $\mathrm{We}^{1/2}$ dependence at fixed groove width, but also exhibits an additional groove-width dependence not contained in \cref{eq:smax-scaling}. 
We quantify this using
\begin{equation}\label{eq:ls-fit-sa-form}
    \frac{S_{\max}}{S_0}
    =
    b + C\, (W/R_0)^p\, \mathrm{We}^{1/2},
\end{equation}
fit on the same in-regime data set. 
The resulting coefficients, $C=0.11$, $b=0.76$, and $p=0.94$, give $\mathrm{RMSE}=0.01$ and $R^2=0.99$. 
Thus the inertio-capillary $\mathrm{We}^{1/2}$ scaling is well reproduced, but with a geometry-dependent prefactor.

The maximum spread length shows the same separation between Weber-number and groove-width effects.
Per-groove least-squares fits proportional to $\mathrm{We}^{1/2}$ are excellent across all five widths, with minimum $R^2=0.9992$, confirming the predicted Weber-number dependence at fixed $W/R_0$.
The groove-width dependence, however, does not collapse onto a single power law over the full range studied. 
The three narrowest grooves are approximately described by a power law in $W/R_0$, but with an exponent closer to $p=-1.63$ than the predicted $p=-1$, while the two widest grooves give nearly coincident values of $\ell_{\max}$ despite continued growth in $S_{\max}$. 
The geometric data therefore preserve the inertio-capillary Weber-number scaling, but not the groove-width dependence implied by the geometric closure employed in the derivations of \cref{sec:groove:scaling}.

\begin{figure}
    \centering
    \includegraphics{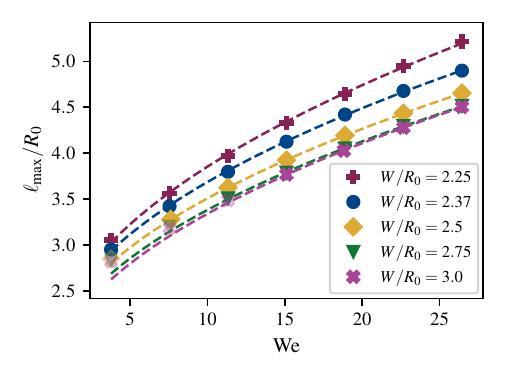}
    \caption{Comparison of $\ell_{\max}$ simulation data with least-square fits of the form $\mathrm{We}^{1/2}$ as predicted by \cref{eq:length-We-scaling}. 
    All grooves show good agreement individually, although the groove-width ordering differs from the predictions of \cref{eq:length-We-scaling}.
    In particular the $\ell_{\max}$ values for the two widest grooves are nearly coincident.
    Faded markers indicate data with $N < 1.20$, excluded from the fits.}
    \label{fig:lmax-scaling}
\end{figure}

Three results therefore emerge with different degrees of robustness. 
First, the blob-decomposition scaling $\tau_c/\tau_0 \sim N^{-1/2}$ of \cref{eq:tc-scaling} is supported directly by \cref{fig:tc-scaling}: data from all five groove widths collapse onto a single trend when plotted against the measured value of $N=\ell_{\max}/W$.
Second, the Weber-number dependence $\tau_c/\tau_0 \sim \mathrm{We}^{-1/4}$ is supported by the imposed-parameter fit in \cref{fig:tc-we-scaling-collapse}, consistent with the inertio-capillary energy-conversion argument underlying the model. 
Third, the groove-width dependence is directionally correct and systematically ordered: at fixed $\mathrm{We}$, wider grooves produce shorter contact-times, as predicted by \cref{eq:tc-We-scaling}. 
This monotonic ordering is non-trivial, and it shows that $W$ remains the ordering parameter, although the magnitude of its influence is under-predicted.
The same pattern appears in the maximum-surface-area and maximum-spread-length data: the $\mathrm{We}^{1/2}$ scaling is reproduced cleanly, while the groove-width dependence is more complex.
We defer further interpretation of the width-exponent discrepancy to \cref{sec:groove:discussion}, noting here that the error stems from the geometric closure, and does not affect the measured-$N$ collapse.

The companion prediction \cref{eq:ta-scaling} for axial retraction is more difficult to test cleanly. 
One issue is that there is no analogue of the contact-time with sharply defined impact and takeoff events that can be used to define $\tau_\parallel$ unambiguously. 
The most reliable proxy is the time between the first and second peaks in the axial kinetic energy component $K_y$, marked (d) and (i) in \cref{fig:droplet-evolution}, which correspond approximately to peak axial expansion and recoil, and therefore to $\tau_\parallel/2$.
Even this measure is imperfect, since the precise extrema of the axial mode may be obscured by higher-order oscillations and harmonics excited during impact.
In addition, once the droplet detaches from the groove, and the geometric distinction between axial and transverse directions is no longer imposed by the boundary, the post-detachment oscillation relaxes toward the radially symmetric Rayleigh modes of a free drop, preventing reliable measurement over multiple periods.
Another issue is that the usable measurement window is narrow. 
At the low-Weber-number, wide-groove end, the data approach the previously identified weak-confinement limit, $N<1.2$, where the blob construction is no longer reliable. 
At the high-Weber-number, narrow-groove end, Rayleigh--Plateau breakup of the elongated droplet can terminate the oscillation before the second peak in $K_y$ occurs.

\begin{figure}
    \centering
    \includegraphics{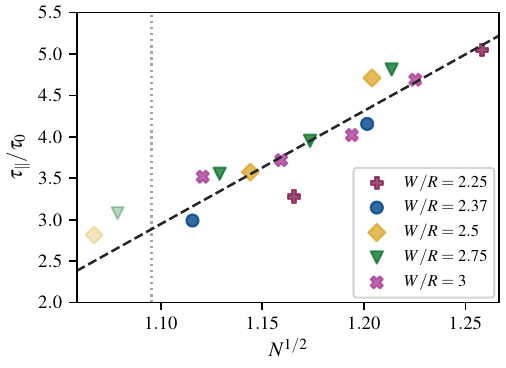}
    \caption{Axial-mode period $\tau_\parallel/\tau_0$, estimated by doubling the interval between successive peaks in the axial kinetic energy component $K_y$, as a function of $N^{1/2}$.
    The dashed black line shows a linear fit. 
    The dotted grey line indicates the $N=1.2$ lower-bound for the blob model, and the faded markers below this are excluded from the least squares fit.
    The data are consistent with the $N^{1/2}$ scaling predicted by \cref{eq:ta-scaling}. 
    Data are restricted at high Weber numbers because Rayleigh--Plateau breakup terminates the oscillation before a second peak can be measured.}
    \label{fig:ta-scaling}
\end{figure}

Despite these caveats, $\tau_\parallel$ shows a clear dependence on $N$, with an approximately linear trend in $N^{1/2}$ consistent with \cref{eq:ta-scaling}.
The measured $\tau_\parallel$ increases monotonically with both $\mathrm{We}$ and $N$ for each groove geometry, and varies by roughly a factor of two across all measurable, in-regime simulations.
The fit is shown in \cref{fig:ta-scaling}. 
Although the prefactor is not tightly constrained, the predicted functional form is supported.

The axial-mode data therefore provide a complementary test of the blob model. 
Imprinted by the confinement of the groove, this time scale depends on the same blob number $N$ that controls the contact-time, and varies oppositely from $\tau_\perp$.
It is therefore distinct from a free-drop Rayleigh oscillation, which would be independent of $N$, and which the measured $\tau_\parallel$ exceeds across all cases. 
Thus both \cref{eq:tc-scaling} and \cref{eq:ta-scaling}, together with their reciprocal relation \cref{eq:mode-product}, are reflected in the simulations despite the greater uncertainty in extracting the axial mode.

\subsection{\label{sec:groove:wetting}Droplet impacts on a wetting groove}

All groove simulations discussed so far used the fully non-wetting boundary condition, $\theta_c = 180^\circ$. 
To probe the effect of finite surface affinity, we repeat the study of \cref{sec:groove:nonwetting} for the same groove widths, $W/R_0 \in \{2.25,2.37,2.5,2.75,3.0\}$, using the calibrated wetting boundary condition of \cref{sec:setup}, for which the static contact angle is $\theta_c \approx 165^\circ$. 
The Weber-number range is extended slightly at the upper end to $\mathrm{We}\in[3,38]$.

\begin{figure}
    \centering
    \includegraphics{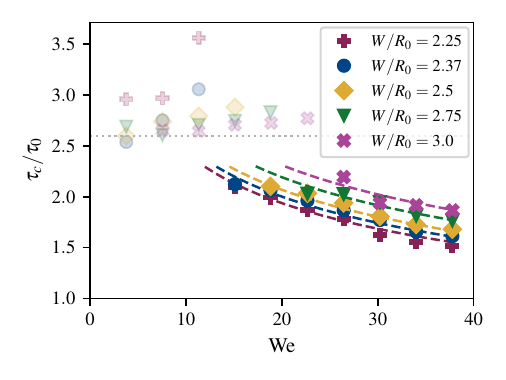}
    \caption{contact-time as a function of Weber number for wetting grooves.
    The data separate into two branches divided by a groove-width-dependent transition Weber number $\mathrm{We}^\star$. 
    Dashed curves show a least-squares fit to the post-transition branch, $\mathrm{We}>\mathrm{We}^\star$, and faded markers indicate data excluded from the fit.
    The dotted gray line is the $\tau_c / \tau_0 = 2.6$ flat plate limit.}
    \label{fig:wet_tc_we}
\end{figure}

\begin{figure}
    \centering
    \includegraphics{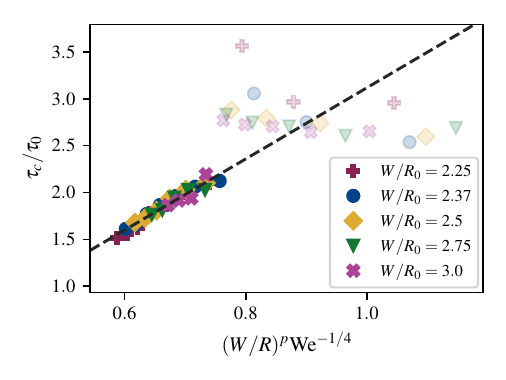}
    \caption{Contact-time $\tau_c/\tau_0$ plotted against $(W/R_0)^p\,\mathrm{We}^{-1/4}$ for the wetting-groove cases. 
    Post-transition impacts, $\mathrm{We}>\mathrm{We}^\star$, collapse onto a common trend, whereas pre-transition impacts follow a distinct branch.
    Faded markers indicate data points exlcluded from the least squares fit.}
    \label{fig:wet_collapse_N}
\end{figure}

The resulting contact-times are shown in \cref{fig:wet_tc_we}. 
Two changes relative to the non-wetting grooves are apparent.
First, finite wall affinity increases the contact-time over the full Weber-number range.
This is consistent with the flat-plate calibration in \cref{sec:validation}, where changing from $\theta_c=180^\circ$ to $\theta_c\approx165^\circ$ increases the contact-time from $\tau_c/\tau_0\approx2.19$ to $\tau_c/\tau_0\approx2.6$. 
Second, the response is no longer monotonic. 
For each groove width, $\tau_c$ first increases with $\mathrm{We}$, reaches a groove-dependent threshold $\mathrm{We}^\star$, then drops sharply onto a second branch that decreases with $\mathrm{We}$, similar to the non-wetting trend but shifted to larger contact-times.

We fit only this post-transition branch, $\mathrm{We}>\mathrm{We}^\star$, using the same procedure and functional form as in \cref{sec:groove:nonwetting}.
The resulting parameters are $C=3.80$, $b=-0.68$, and $p=0.46$ with $\mathrm{RMSE}=0.04$ and $R^2=0.95$. 
The data collapse onto an approximately linear trend in $(W/R_0)^p\,\mathrm{We}^{-1/4}$, as shown in \cref{fig:wet_collapse_N}.
The fitted width exponent is much smaller than in the non-wetting case, where $p\approx1.51$. 
Thus wettability affects not only the contact-time offset, but also the apparent groove-width dependence of the fast branch, reducing the exponent by approximately one power of $W/R_0$.

The wetting-groove data therefore introduce two issues that must be explained: the reduction of the groove-width exponent from super-linear to sub-linear, and the appearance of a sharp transition between a low-Weber-number branch with increasing contact-time and a high-Weber-number branch with decreasing contact-time.

\subsubsection{\label{sec:groove:wetting:modfast}The groove-width exponent}

\begin{figure}
    \centering
    \includegraphics{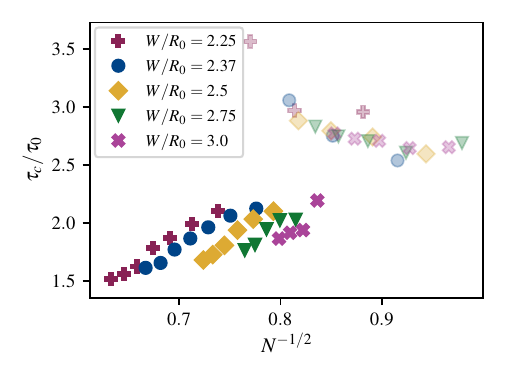}
    \caption{Post-transition wetting data, $\mathrm{We}>\mathrm{We}^\star$, plotted against $N^{-1/2}$. 
    Like the non-wetting data in \cref{fig:tc-scaling}, the wetting data are approximately linear for each groove width, however they do not collapse onto a single line, with the curves offset from one another ordered by groove width. 
    This indicates an additional groove-width dependence associated with finite wall affinity. 
    The transition between the axial and transverse branches also does not occur at a single value of $N$, indicating that droplet deformation alone is not sufficient to determine the mode selection.
    Faded markers indicate data on the low-$\mathrm{We}$ number branch, not under consideration.}
    \label{fig:wetting_n_noncollapse}
\end{figure}

For the wetting boundary cases, the change in the fitted groove-width exponent is already apparent in the measured-$N$ representation. 
As shown in \cref{fig:wetting_n_noncollapse}, the data for each groove width retain an approximately linear dependence on $N^{-1/2}$, but no longer collapse onto a common line.
The per-width fits have similar slopes but also systematic offsets, indicating that a residual dependence on $W$, not accounted for by the idealized blob model, remains.

This differs fundamentally from the non-wetting case. 
There, the measured-$N$ collapse was successful, and the anomalous fitted exponent $p\approx1.51$ in \cref{eq:ls-fit-form} arose from the geometric closure of \cref{eq:N-scaling} used to eliminate $N$ in favor of the imposed parameters $W$ and $\mathrm{We}$.
Under finite wall affinity, the deviation appears at the earlier step.
The additional width dependence is therefore not purely an artifact of the limited geometric representation, but reflects a wetting-sensitive contribution to the rebound dynamics.

A plausible origin is the interaction between the liquid and the groove walls during retraction and detachment. 
Contact-line motion and liquid--solid affinity which resists separation from the groove walls both provide additional pathways for dissipation.
These effects depend on the evolving liquid--solid contact geometry and can therefore introduce an explicit dependence on $W$ that is not contained in the inertio-capillary blob model.

The persistence of this residual width dependence does not imply a breakdown of the leading Weber-number scaling.
Fits of $S_{\max}$ and $\ell_{\max}$ in the wetting cases show the same qualitative behavior as in \cref{sec:groove:nonwetting}: the predicted $\mathrm{We}^{1/2}$ dependence is retained at fixed groove width, while the prefactors remain width dependent. 
Wetting therefore alters the morphology- and wall-interaction-sensitive dependence on $W$ without substantially changing the global inertio-capillary dependence on $\mathrm{We}$.

\subsubsection{\label{sec:groove:wetting:bimodal}The mode-selection transition}

\begin{figure*}
    \centering
    \includegraphics[width=\textwidth]{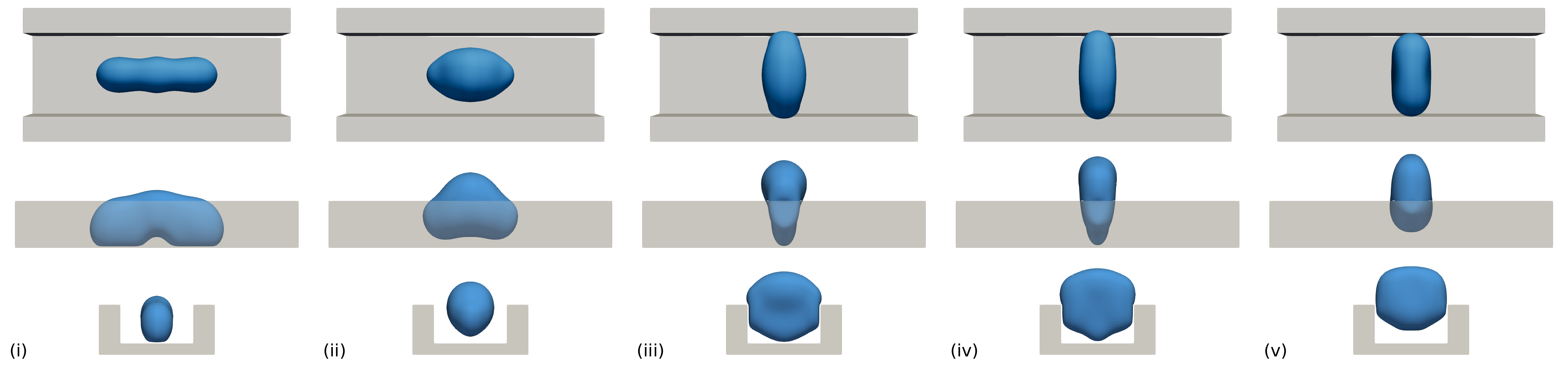}
    \caption{Evolution of a droplet with initial Weber number $\mathrm{We}=15.1$ rebounding from a groove with $W/R_0=2.5$ as seen from the top (top), side (middle) and end (bottom). 
    This case lies slightly below $\mathrm{We}^{\star}$, thus the contact-time is controlled by the axial mode $\tau_\parallel$.
    \textbf{(i)} The droplet retracts transversely and rising from the groove surface.
    \textbf{(ii)} The droplet temporarily departs the surface while contracting axially.
    \textbf{(iii)} Subsequent radial expansion in the transverse--vertical plane re-establishes contact with the groove bottom and walls.
    \textbf{(iv)} The droplet begins to separate a second time, forming a liquid neck near the departure point.
    \textbf{(v)} The droplet fully detaches from the groove bottom.
    In lower $\mathrm{We}$ cases, the full separation and re-attachment evident here does not occur, although the mechanism of prolonged contact-time is the same.
    The necking visible in \textbf{(iv)} is caused by wetting affinity between the liquid phase and groove surface, and is therefore absent in the non-wetting cases.}
    \label{fig:droplet_evolution_near_cliff}
\end{figure*}

Investigation of the evolution of the droplet shape during rebound offers insight into the appearance of two branches in \cref{fig:wet_tc_we,fig:wet_collapse_N}.
As in the non-wetting case, the groove stretches the droplet axially, so that it detaches in an elongated configuration. 
Capillarity then drives an axial recoil as the droplet rises, which volume conservation dictates is accompanied by a radial expansion in the transverse--vertical plane, about the droplet's rising axis.
This evolution is shown in \cref{fig:droplet_evolution_near_cliff}.

\begin{figure}
    \centering
    \includegraphics{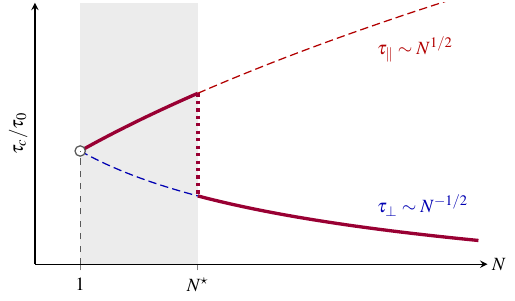}
    \caption{Schematic illustration of the mode-selection transition for fixed groove width. 
    At $N=1$, the groove does not constrain the droplet and the axial and transverse time scales coincide. 
    For intermediate $N$, the slow axial mode controls the measured contact-time. 
    At $N^\star$, which depends on $W/R_0$, the controlling mode switches from the axial mode to the transverse recoil mode.}
    \label{fig:cliff_schematic}
\end{figure}

At low Weber numbers, i.e. $\mathrm{We} < \mathrm{We}^\star$, this transverse--axial expansion occurs before the droplet has risen far from the groove, or even fully detached.
The vertical expansion therefore either prolongs or re-establishes contact. 
In this regime, $\tau_c$ is governed by the slow axial mode $\tau_c \sim \tau_\parallel$ which scales according to \cref{eq:ta-We-scaling}, thus increasing with Weber number. 
For $\mathrm{We}>\mathrm{We}^\star$, the droplet rises rapidly enough that the transverse--vertical expansion, at its maximum extent, no longer reaches the substrate. 
The contact-time is then set by the faster transverse recoil $\tau_c \sim \tau_\perp$ described by \cref{eq:tc-We-scaling}, giving the decreasing branch observed at larger Weber number, and a sharp transition between the two. 
The resulting mode-selection picture is summarized schematically in \cref{fig:cliff_schematic}.

To quantify the cross-over point, we model the transition as a race between the droplet's rebound from the substrate and its axial recoil. 
The relevant quantities are the rebound velocity $U_r$, the axial time scale $\tau_\parallel$, and the characteristic vertical extent $R_v$ reached during the transverse--vertical expansion. 
We employ the scalings
\begin{subequations}
    \begin{align}
        U_r &\sim \mathrm{We}^{1/2}, \label{eq:ureb}\\
        \tau_\parallel/\tau_0  &\sim N^{1/2}, \label{eq:taua}\\
        R_v/R_0        &\sim \mathrm{We}^{1/4}, \label{eq:rv}
    \end{align}
\end{subequations}
which imply the following assumptions.
First, \cref{eq:ureb} assumes that the rebound velocity remains proportional to the impact velocity, $U_r = e U_0$, with restitution factor $e=\mathcal{O}(1)$. 
\cref{eq:taua} follows the axial-mode scaling derived in \cref{sec:groove:scaling}. 
Finally with \cref{eq:rv} we propose that the transverse--vertical expansion after axial collapse follows the same energy-limited scaling as the maximum spreading radius on a flat plate.

\begin{figure}
    \centering
    \includegraphics{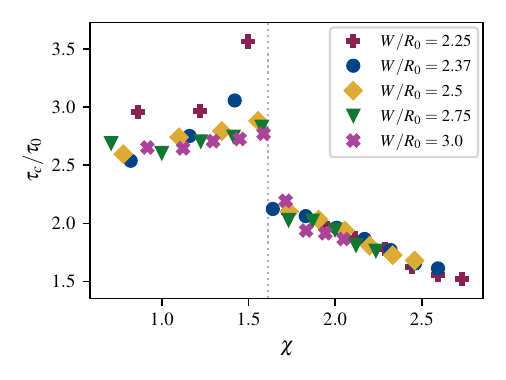}
    \caption{Contact-time plotted against the mode-selection parameter $\chi\sim(W/R_0)^{-1}\,\mathrm{We}^{1/2}$. The two regimes separate near $\chi^\star\approx1.61$.}
    \label{fig:tc-v-chi}
\end{figure}

\begin{figure}
    \centering
    \includegraphics{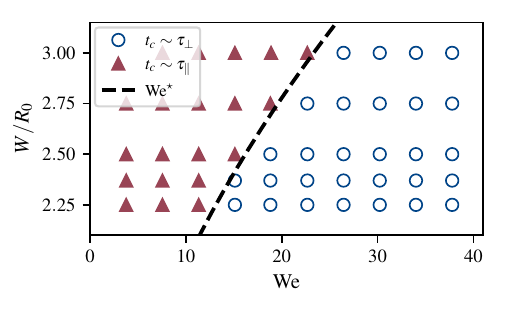}
    \caption{Comparison of simulation data with the estimated transition Weber number $\mathrm{We}^\star$.
    Impacts with responses on the slow branch associated with $\tau_\parallel$ are marked with red triangles; those associated with the fast branch $\tau_\perp$ are blue circles.
    The dashed black line is the transition criterion predicted by $\mathrm{We}^\star = (\chi^\star W / R_0)^2$ with $\chi^\star = 1.61$ from \cref{fig:tc-v-chi}.
    The transition criterion cleanly partitions the data.}
    \label{fig:w-v-weber-mode-selection}
\end{figure}

The transition is then controlled by the ratio of the rebound speed $U_r$ to the distance the droplet must clear $R_v$ and the time it has to do so $\tau_\parallel$:
\begin{equation}\label{eq:chi}
    \chi \equiv \frac{U_r}{R_v/\tau_\parallel}
    \sim \mathrm{We}^{1/4}N^{1/2}
    \sim \frac{\mathrm{We}^{1/2}}{W/R_0},
\end{equation}
where the final relation follows from \cref{eq:N-scaling}.
Re-contact is suppressed, and the fast transverse mode controls the contact-time, once $\chi$ exceeds a constant threshold $\chi^\star$. 
For a constant value of $\chi=\chi^\star$, \cref{eq:chi} may be rearranged to give
\begin{equation}\label{eq:we-star-scaling}
    \mathrm{We}^\star \propto (W/R_0)^2.
\end{equation}
In \cref{fig:tc-v-chi}, plotting $\tau_c$ against $\chi$ separates the two regimes, with the transition occurring near $\chi^\star\approx1.61$. 
The Weber-number sampling is not fine enough to determine $\mathrm{We}^\star$ accurately for each groove width, so \cref{eq:we-star-scaling} cannot be tested quantitatively here.
However, the observed regime separation is consistent with the proposed scaling.
An approximation of $\mathrm{We}^\star$, based on the value of $\chi^\star$ mentioned above, is compared with the observed data in \cref{fig:w-v-weber-mode-selection}, and found to be consistent.

Finally, we note that the existence of a distinct low-$\mathrm{We}$ branch receives some support from Ref.~\onlinecite{chantelotWaterRingbouncingRepellent2018}, which, as mentioned in \cref{sec:validation}, reports that contact-time measurements on wetting grooves were sufficiently scattered to motivate a change of experimental protocol. 
On the slow branch, contact-time is sensitive to the rebound velocity, so that under experimental noise it could plausibly present as scatter. 
The appearance of such protocol-disrupting behavior for a wetting boundary, over the reported range of Weber numbers, is consistent with the present simulations.

\section{\label{sec:groove:discussion}Discussion}

Several qualifications are required in interpreting the simulation results presented here.
The first concerns the degree of agreement that should be expected from an idealized inertio-capillary model.
The blob model interprets a three-dimensional, dissipative droplet motion through an idealized one-dimensional mass--spring analogy, retaining only a small number of effective geometric and inertial degrees of freedom.
In simulations, however, impact energy is also distributed into higher-order oscillations, internal circulation, viscous dissipation, interfacial deformation, and, in some cases, capillary breakup.
These effects are outside the blob model description, and consequently exact collapse onto the predicted scalings cannot be expected.
Finite wall affinity introduces additional sources of deviation through contact-line motion and liquid--solid adhesion.
This is consistent with the larger scatter and reduced fit quality observed for wetting groove impacts. 
The numerical data should therefore be interpreted as corroborating the scaling laws derived from physical arguments, rather than as an exact validation of the idealized model.

The second qualification concerns the quantitative resolution of the scaling analysis.
The accessible parameter range is limited by the finite physical window in which the groove-imposed blob description remains applicable.
Low-Weber-number impacts on wide grooves approach the flat-plate limit, while high-Weber-number impacts on narrow grooves produce elongated, thinned droplets susceptible to Rayleigh--Plateau breakup.
The current data set spans five groove widths $W/R_0\in[2.25,3.0]$ over a relatively narrow band.
This is sufficient to identify systematic trends but limits the precision with which exponents, prefactors, and crossover points can be determined.
The transition from a super-linear width dependence for contact-time in the non-wetting case to a sub-linear dependence in the wetting case is therefore interpreted as robust, whereas the numerical value of the exponent difference should be regarded as indicative rather than definitive.
A similar interpretation should be made for the width dependency exponents applied to the surface area and spread length scalings in \cref{sec:groove:nonwetting} --- both deviate meaningfully from the predicted behavior, but the exact relationship is not well established.

The results in \cref{sec:groove:nonwetting,sec:groove:wetting} reveal a consistent hierarchy in the predictive accuracy of the model.
Across both wetting and non-wetting cases, the Weber-number exponents of $S_{\max}$, $\ell_{\max}$, and $\tau_c$ are reproduced cleanly, whereas their dependence on $W/R_0$ is captured only approximately. 
The groove-width trends remain directionally correct, but the fitted exponents differ from the idealized predictions.
While the wetting case appears to introduce new, groove-width dependent dynamics, the deviations in the non-wetting case stem from inaccuracy of the geometric closure in representing the spread morphology of the droplet.

This inaccuracy stems from a core limitation of the blob model: the form of the spread droplet is insufficiently constrained.
The model assumes that at the point of maximum spreading, the droplet roughly takes on the form of a rectangular prism, defined by four variables $W$, $\ell_{\max}$, $h$, and $S_{\max}$.
However, the relationship between these variables is constrained by only two relationships: volume conservation $V \sim W\, h\, \ell_{\max}$, and the area closure $S_{\max} \sim W\, \ell_{\max}$.
To eliminate the remaining freedom, the imposed-parameter scalings of \cref{sec:groove:scaling} propose the simplest closure: the inertio-capillary result $S_{\max}\sim\mathrm{We}^{1/2}$ of \cref{eq:smax-scaling}. 
What survives this simplification is anything determined by global energy balance.
What does not survive are predictions requiring a commitment to how large $S_{\max}$ is, and how it is distributed across the geometric features.
The droplet during impact is free to adjust the maximum surface area, as well as redistribute it by adjusting $h$ and $\ell_{\max}$, not to mention deviate from the prismatic assumption on spread morphology altogether.
Even the width $W$ is not truly fixed, as the droplet in some cases may spread vertically up the sides of the groove walls as seen in \cref{fig:droplet-evolution}.
However, the simplified geometric closure adopted in \cref{eq:smax-scaling} succeeds in providing a directionally correct, although not exact, scaling for the effect of groove width.

The difficulties in interpreting the simulation data described above, prevent us from drawing any further quantitative conclusions about how wetting modifies the contact-time scaling, although several features of the data seem suggestive.
In particular, the fast branch appears to collapse linearly with $\chi$ in \cref{fig:tc-v-chi}, which implies a $\tau_c / \tau_0 \sim \mathrm{We}^{-1/2}$ relationship.
This change in energy scaling does not necessarily contradict the robustness of the inertio-capillary picture across the two wetting conditions just described. 
Finite wall affinity may modify release dynamics while leaving the core energy exchange intact.
The wetting data, however, exhibit appreciable scatter, and the two branch structure reduces the number of applicable points.
Consequently, a clear adjudication between a $\mathrm{We}^{-1/2}$ and $\mathrm{We}^{-1/4}$ form is not possible.
Additionally, as only a single wetting condition has been investigated, any observed dependency cannot be extrapolated beyond the particular boundary condition studied.
Most importantly, though, there is no a priori reason to expect the effects introduced by finite wall affinity to follow simple power laws in $W$ or $\mathrm{We}$.
The inertio-capillary blob model is based on an idealized reversible exchange between kinetic and surface energy, whereas wall affinity introduces geometry-dependent dissipative processes.
In many cases, dissipative boundary effects added to otherwise inertial flow models introduce new sensitivities that cannot be captured by the same low-dimensional scaling structures.
The present data do not cleanly distinguish between a modified scaling law and a lower-order correction to the blob-model dynamics.
We therefore report these trends as observations, noting that systematic interpretation would require data spanning a range of contact angles.

The robustness of the maximum surface area scaling, and inertio-capillary energy exchange, motivates the form of $\chi$, through the approximation $R_v/R_0\sim\mathrm{We}^{1/4}$ in \cref{eq:rv}.
This estimate is adopted by analogy with the flat-plate maximum-spread radius in \cref{eq:max_spreading_radius}, which also underlies the area scaling in \cref{eq:smax-scaling}, rather than being derived directly from the axial-collapse dynamics. 
The vertical re-expansion following axial recoil need not, in principle, share the same exponent as the primary transverse spread.
However, estimating its magnitude using the same inertio-capillary length scale is consistent with the view that rebound remains organized by continued exchange between kinetic and surface energy.
The collapse of the transition near a single value $\chi^\star\approx1.61$ in \cref{fig:tc-v-chi} supports this approximation. 
If the $\mathrm{We}^{1/4}$ length-scale estimate were substantially inappropriate, residual groove-width ordering would be expected near the transition, which is not evident in the present data.

Finally, the numerical value of $\chi^\star$ should be regarded as specific to the boundary condition studied here. 
Since only one wetting boundary condition was tested, the dependence of $\chi^\star$ on contact angle cannot be determined from the present data, although the expected direction of the effect can be inferred. 
As wall affinity vanishes, the dissipative reduction of the rebound velocity $U_r$ should weaken, shifting the transition to lower $\chi$.
Consistent with this interpretation, the non-wetting simulations in \cref{sec:groove:nonwetting} exhibit a single monotonic branch over the corresponding range of $\chi$, with no mode-selection transition.
Thus any non-wetting threshold appears to lie below the Weber-number range sampled here.

An examination of the limiting behavior of $\chi$ suggests that a resolvable value in the non-wetting limit may not exist.
From \cref{eq:chi} and \cref{eq:N-scaling}, we can state that
\begin{equation}
    \frac{\chi^4}{N^2} \sim \mathrm{We}.
\end{equation}
As the problem definition requires $N \geq 1$, small values of $\chi$ imply vanishing Weber number, where the pure inertio-capillary assumption becomes invalid for a real droplet as viscous effects increasingly dominate.
Thus, in the limit of vanishing wall affinity, $\chi^\star$ may be outside the range of validity of the present mode-selection model, although this cannot be resolved without additional low-$\mathrm{We}$ simulations.

\section{Conclusions}

A non-ideal, entropic, multi-relaxation-time lattice Boltzmann method was used to study droplet impacts on superhydrophobic grooves.
Following validation against experimental measurements for superhydrophobic surfaces and non-wetting Leidenfrost grooves, simulations were used to examine how groove geometry breaks the radial symmetry of rebound and separates the impact into transverse and axial dynamics.

The principal contribution is an extension of the blob model of Ref.~\onlinecite{chantelotWaterRingbouncingRepellent2018} to a two-mode description of droplet--groove impact. 
The groove imposes reciprocal transverse and axial inertio-capillary time scales, described by \cref{eq:tc-scaling,eq:ta-scaling} and related by \cref{eq:mode-product}.
Simulations across five groove widths with non-wetting boundaries confirm this framework: contact-times collapse onto the predicted transverse scaling of \cref{eq:tc-scaling}, and the complementary axial mode, though harder to extract, follows its anticipated trend. 
For imposed-parameter scaling proposed in \cref{eq:tc-We-scaling}, the predicted Weber number dependence is recovered, but the groove-width dependency is found to be super-linear. 
This is attributable to limitation of the rectangular-prism closure in describing the geometry of the true, spread droplet.

Extending the study to finite surface affinity reveals a non-monotonic, two-branch contact-time response: contact-time increases with Weber number on the low-$\mathrm{We}$ branch, decreases on the high-$\mathrm{We}$ branch, and switches sharply at a groove-width-dependent $\mathrm{We}^\star$.
The two-mode model receives further support from its use in interpreting this behavior, with the two branches each associated with one of the two primary modes. 
The transition between branches is modeled as a competition between wetting-slowed rebound and axial recoil, with the resulting groove-width-dependent transition points collapse onto the single mode-selection parameter $\chi$ of \cref{eq:chi}.

Predictions tied to global energy balance are shown to be reproduced cleanly across both wetting and non-wetting cases. 
Predictions tied more directly to droplet morphology are shown to be directionally correct, although quantitatively inexact.
The reasons for this are explored in \cref{sec:groove:discussion}.

The main result is therefore a unified interpretation of groove-mediated rebound in terms of two inversely related inertio-capillary modes.
The simulations support the central physical picture: groove geometry breaks the degeneracy of flat-plate rebound, imposes distinct transverse and axial time scales, and allows the observable contact-time to be controlled through mode selection.
This distinguishes anisotropic confinement from strategies that primarily shorten an existing rebound pathway by reducing the active mass or retraction length.
The fixed transverse length therefore changes not only the rebound time, but the selection mechanism that determines it.
These findings clarify how geometric confinement and wetting effects modify droplet rebound on superhydrophobic surfaces, with implications for textured surfaces used in anti-icing, self-cleaning, and droplet transport.

\section*{Acknowledgments}
This work was supported by European Research Council (ERC) Advanced Grant No. 834763-PonD and by the Swiss National Science Foundation (SNSF) Grant 200021-228065.
Computational resources at the Swiss National Super Computing Center (CSCS) were provided under Grants No. s1286 and s1327. 

\section*{Author Declarations}

\noindent\textbf{Conflict of Interest}\\
The authors have no conflicts to disclose.

\noindent\textbf{Ethical approval}\\
The work presented here by the authors did not require ethics approval or consent to participate.

\noindent\textbf{Author Contributions}\\
\textbf{M. Feinberg}: Conceptualization (equal); Formal analysis (lead); Investigation (lead); Methodology (supporting); Software (lead); Visualization (lead); Writing -- original draft (lead); Writing -- review \& editing (equal).
\textbf{S. A. Hosseini}: Conceptualization (equal); Formal analysis (supporting); Methodology (lead); Software (supporting); Supervision (equal); Writing -- review \& editing (equal).
\textbf{I. V. Karlin}: Conceptualization (equal); Funding acquisition (lead); Project administration (lead); Resources (lead); Supervision (equal); Writing -- review \& editing (equal).

\noindent\textbf{Data availability}\\
The data that support the findings of this study are available from the corresponding author upon reasonable request.

\appendix

\section{\label{app:surfacetension} Measurement of surface tension}

Although the surface tension coefficient $\gamma$ does not appear explicitly as an input to the LB evolution equation, it is required to evaluate both the inertio-capillary time scale and the Weber number. 
It was therefore measured from the Laplace pressure of static liquid columns with known curvature,
\begin{equation}
    \Delta P = \frac{\gamma}{R_e},
\end{equation}
where $\Delta P$ is the pressure jump across the interface, and $R_e$ is the equimolar radius defined by the Gibbs dividing surface.\cite{gibbsEquilibriumHeterogeneousSubstances1878} 
Simulations were performed for cylindrical liquid columns of several initial radii using the same thermodynamic state and transport properties as in the impact simulations. 
After convergence of the density field, $\Delta P$ was computed using \cref{eq:vdw} from bulk liquid and vapor densities, while $R_e$ was evaluated from
\begin{equation}
    R_e = \sqrt{\frac{\int_0^{\infty}2(\rho(r) - \rho_v)\,r\, dr}{(\rho_l - \rho_v)}}.
\end{equation}

Both quantities are subject to numerical uncertainties: $\Delta P$ is sensitive to the choice of bulk sampling points, particularly in the liquid phase, where the repulsive denominator $1-b\rho$ in \cref{eq:vdw} makes the pressure response stiff, so that small variations in $\rho$ can produce appreciable changes in $P$.
Meanwhile, $R_e$ is affected by the imperfect representation of a curved interface on a Cartesian lattice, and consequently the value depends on the exact measurement approach.\cite{hosseiniLatticeBoltzmannNonideal}

\begin{figure}
    \includegraphics{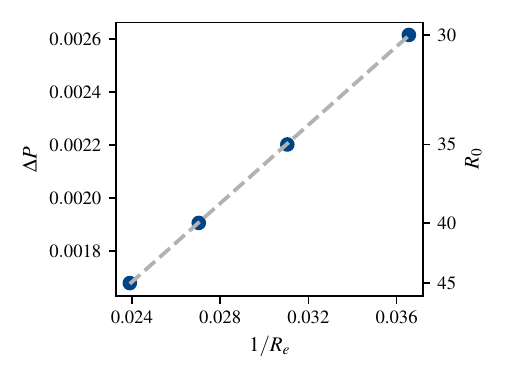}
    \caption{Pressure jump as a function of inverse equimolar radius for static liquid columns at $T/T_c = 0.3936$. 
    The dashed line is a least-squares fit, whose slope gives the surface tension.}
    \label{fig:surface_tension_measurement}
\end{figure}

The effective surface tension was obtained from a least-squares fit of $\Delta P$ against $1/R_e$ across the four column simulations, with the fitted slope identified as $\gamma$. 
This gives $\gamma \approx 0.0704$.
The recovered linear dependence confirms that the model reproduces the expected relation between pressure jump and curvature; see \cref{fig:surface_tension_measurement}.
Variations in $\gamma$ arising from alternative measurement choices do not affect the scaling behaviors discussed in the main text, nor appreciably alter conclusions drawn from comparison with the experimental data.

The converged density ratio is $\rho_l/\rho_v \approx 578$, slightly below the target value of 600 due primarily to vapor-phase drift associated with the under-resolved diffuse interface. 
The interface width is $w = 4.56\,\delta r$, measured as
\begin{equation}
    w = \frac{\rho_l - \rho_v}{\max |\nabla \rho|}.
\end{equation}
This small departure from the Maxwell-construction prediction is not expected to affect the impact dynamics considered here.

Spurious currents are weak, non-physical velocities which arise near curved interfaces due to small discretization errors in the balance between thermodynamic pressure gradients and capillary forces. 
They are a practically unavoidable feature of diffuse interface methods, and it is necessary to ensure scale separation between these spurious velocities and the characteristic velocities of the flow.
To assess these, we measure the peak velocity in the smallest converged liquid column, which has a curvature comparable to that of the three-dimensional droplet.
The peak spurious-current magnitude observed is $|\bm{u}_{\text{spurious}}| \approx 1.5 \times 10^{-3}$, roughly a factor of fourteen below the slowest impact velocity used in this study. 
Additionally, peak spurious velocities occur in the vapor phase, further reducing their dynamical influence given the high density ratio.

\begin{figure}
    \includegraphics[width=\linewidth]{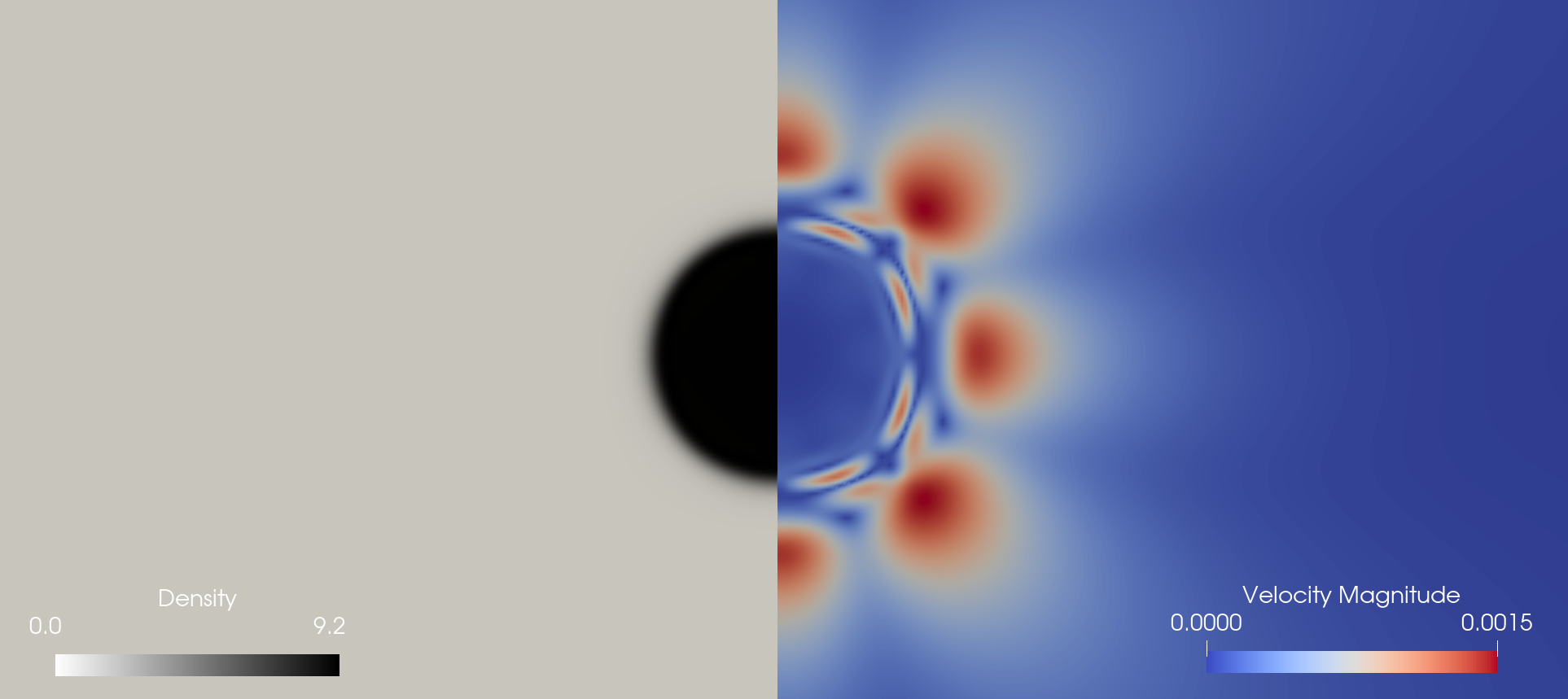}
    \caption{\label{fig:spurious_currents} \textbf{Left:} density field around a static droplet. \textbf{Right:} velocity magnitude of the corresponding spurious currents.}
\end{figure}

\section{\label{app:wetting}Wetting boundary condition}

The fictitious, wall-density wetting BC employed requires calibration of $\alpha_w$ to the desired contact angle.
Characterization of this boundary condition with the present free-energy formulation has been reported in Ref.~\onlinecite{hosseiniConsistentLatticeBoltzmann2022}, so here we restrict attention to the two operating points used in the impact simulations: $\rho_w = 0.695$ for the wetting (high-contact-angle SHS) case, and $\rho_w = \rho_v = 0.015$ for the Leidenfrost (non-wetting) case.

\begin{figure}
    \includegraphics[width=\linewidth]{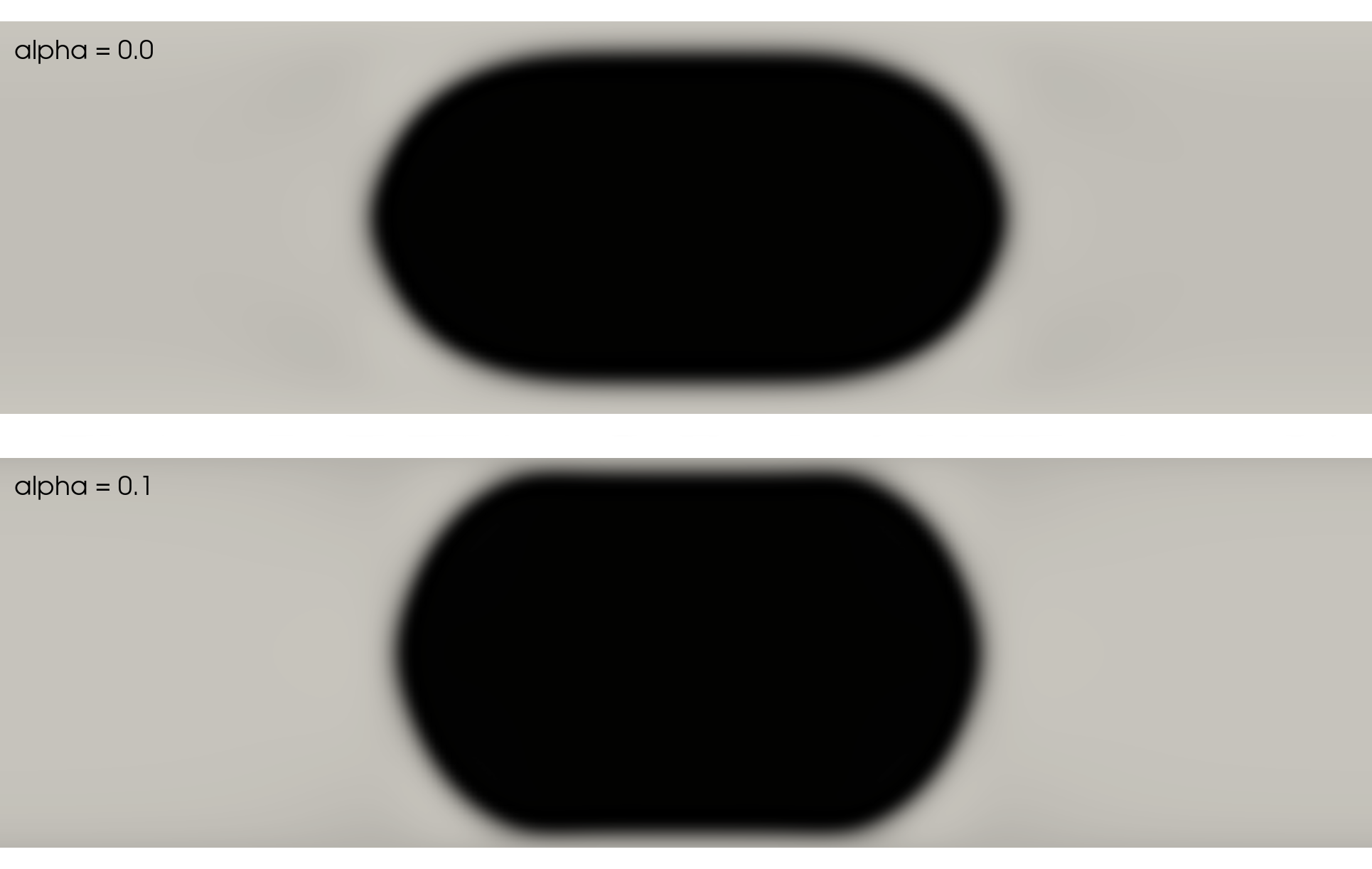}
    \caption{\label{fig:wetting_contact_angle}Converged density fields for a 2D static droplet on a flat substrate at the two operating points. \textbf{Top:} $\rho_w = \rho_v=0.015$, giving $\theta_c \approx 180^\circ$ with no substrate attachment. \textbf{Bottom:} $\rho_w = 0.695$, giving $\theta_c \approx 165^\circ$.}
\end{figure}

Contact angle is assessed from simulations of a liquid column between two flat boundaries. 
After equilibration, $\theta_c$ may be estimated either from the pressure jump across the interface using the Young--Laplace relation
\begin{eqnarray}
    \Delta P = \frac{2\gamma\cos\theta_c}{H},
\end{eqnarray}
where $H$ is the channel height, or by direct measurement of the density field.
The pressure-based estimate is sensitive to small errors in the measured bulk densities as discussed in \cref{app:surfacetension}, which are amplified when converting $\Delta P$ to $\theta_c$ near the wetting limits.
We therefore use direct density-field measurements for calibration, fitting circular arcs to the free interfaces while varying the threshold density and corner exclusion to estimate uncertainty. 
This gives $\theta_c = 165^\circ \pm 3^\circ$ for $\rho_w=0.695$. 
For $\rho_w=\rho_v$, the droplet does not attach to the substrate, consistent with a fully non-wetting condition, $\theta_c \approx 180^\circ$.
The two configurations are shown in \cref{fig:wetting_contact_angle}, where the different wetting behavior is visually apparent.

\section{\label{ap:axial-mode-inertial}Robustness of the axial-mode scaling}

The Taylor--Culick derivation of \cref{sec:groove:scaling} provides a specific mechanism for the axial retraction: a capillary force withdraws a rim of increasing mass at the constant Taylor--Culick velocity. 
To demonstrate the robustness of the $N^{1/2}$ scaling to modeling assumptions, we offer an alternative picture: a capillary force retracts a constant fraction of the droplet's mass back toward its center at constant acceleration.

Combining Newton's second law with the classical kinematic equation $x-x_0=\frac{1}{2}at^2$, gives us a generic expression for the time-scale of dynamics dominated by constant acceleration
\begin{equation}\label{eq:general_time_scaling_newton_kinematic}
    t = \left( \frac{m (x-x_0)}{F} \right)^{1/2}.
\end{equation}
We proceed by assuming the mass of the droplet's end experiencing this acceleration is proportional to the total mass of the droplet $m \sim \rho_l R_0^3$, and that the length scale of the retraction $(x-x_0) \sim \ell_{\max}$.

The capillary force driving the recoil can be derived from multiple modeling assumptions.
It can be found from the gradient of the surface energy ($F_\gamma = -\gamma\,dA/dy$), from line tension acting along the perimeter where free surfaces meet the prismatic end-cap of the spread droplet, or from the Young--Laplace pressure integrated over the curved rim that real recoiling ends form.
The three derivations yield the same scaling form if the spread droplet is assumed to have a rectangular prismatic form,
\begin{equation}
    F_{\gamma} \sim \gamma\, W,
\end{equation}
differing only in the leading $\mathcal{O}(1)$ prefactor.
For example, if we assume that the acceleration comes from the force of the upper and lower surfaces pulling on the end cap, then $F_\gamma = 2\,\gamma\,W$, with contributions from the remaining two sides of the end-cap neglected because $h \ll W$.

Inserting our scaling laws for $F_\gamma$, $m$, and $x-x_0$ into \cref{eq:general_time_scaling_newton_kinematic} results in
\begin{equation}\label{eq:tau_a_final}
    \tau_\parallel \sim \left( \frac{\rho R_0^3}{\gamma}\,\frac{\ell_{\max}}{W} \right)^{1/2} 
    = \tau_0\,\left(\frac{\ell_{\max}}{W}\right)^{1/2} = \tau_0\,N^{1/2},
\end{equation}
recovering the same $N^{1/2}$ scaling arrived at by Taylor--Culick in \cref{sec:groove}.

The reason for the agreement is that both arguments respect the same set of underlying assumptions:
\begin{enumerate}
    \item[(i)] the relevant inertial mass scales with the droplet mass,
    \item[(ii)] the retraction proceeds over a length proportional to the 
    maximum spreading length of the droplet, 
    \item[(iii)] the retraction is driven by a capillary force, which takes the form $F_\gamma \sim \gamma L,$ where $L$ is a length scale characteristic of the recoiling end,
    \item[(iv)] the characteristic length $L$ is set by the groove width $W$.
\end{enumerate}

Assumption (i) is explicit in the present derivation, but is implied by the rate of mass accretion in the Taylor--Culick argument.
Assumptions (i) and (ii) commit to the inertia and the recoil distance scaling with the droplet rather than with some sub-structure of it.
Assumption (iii) reflects the dimensions of surface tension as force per unit length. 

Assumption (iv) requires the most justification.
It is arrived at from the core geometric premise of the blob model of the droplet-groove interaction: that the droplet flattens out into a rectangular prismatic form at the moment of maximum spreading.
This places a clear validity window on the scaling.
When the droplet has not spread sufficiently for $\ell_{\max}$ to significantly exceed $W$, whether because the groove is too wide or because the impact energy is too low, the recoiling end is not prismatic and $W$ is not preferentially the dominant length scale of the force.
In this regime the droplet's behavior approaches that of an unrestrained impact on a flat surface, and the $N^{1/2}$ scaling should not be expected to hold.

The agreement between the two derivations demonstrates that the $N^{1/2}$ exponent is not an artifact of the Taylor--Culick mechanism or of any particular kinematic idealization: it follows from assumptions (i)--(iv), which both derivations share. 
The scaling should therefore be expected to hold wherever those four assumptions hold, and to break down in the wide-groove, low-Weber regime where the spread droplet has not formed the requisite rectangular prismatic shape and assumption (iv) is violated.

\bibliography{groove}

\end{document}